%% file: Ex3DSolLam.tex
\documentclass[3p,10pt]{my_elsarticle}
\usepackage{etoolbox}
\newtoggle{submission}
\toggletrue{submission}
\biboptions{longnamesfirst,square,sort,compress}
\usepackage[english]{babel}
\usepackage[latin1]{inputenc}
\usepackage[T1]{fontenc}
\usepackage{graphicx}
\usepackage{enumerate}

\usepackage{amssymb,amsmath}
\usepackage{empheq}
\usepackage{mathtools}
\usepackage{mathrsfs} 
\usepackage{frcursive}
\usepackage{color}
\usepackage{comment}
\usepackage{multirow}
\usepackage{booktabs}
\usepackage{rotating} 
\usepackage{dcolumn}
\usepackage{wasysym}

\usepackage[scaled=0.6]{helvet}

\iftoggle{submission}{}{
  
  \usepackage{pgfplots}
  \usetikzlibrary{pgfplots.groupplots}
  \pgfplotsset{compat=1.7}
  \tikzset{every mark/.append style={scale=0.5}}
  \usepgfplotslibrary{external}
  \tikzexternalize
}

\newcolumntype{d}[1]{D{.}{.}{#1}}

\newcommand{\ppt}{~\permil}
\newcommand{\ppm}{~\text{ppm}}

\everymath{\displaystyle}
\allowdisplaybreaks[4]
\usepackage{hyperref}
\hypersetup{
    unicode=true,          
    pdftoolbar=true,        
    pdfmenubar=true,        
    pdffitwindow=false,     
    pdfstartview={FitH},    
    pdfnewwindow=true,      
    colorlinks=true,       
    linkcolor=red,          
    citecolor=green,        
    filecolor=magenta,      
    urlcolor=cyan           
}
\begin{document}

\begin{frontmatter}
	\title{Exact 3D solution for static and damped harmonic response\\ of simply supported general laminates}
	\author[drive]{A.~Loredo\corref{cor1}}
	\ead{alexandre.loredo@u-bourgogne.fr}
	%
	%
	\cortext[cor1]{Corresponding author}
	%
	\address[drive]{DRIVE, Université de Bourgogne, 49 rue Mlle Bourgeois, 58027 Nevers, France}
	%
%
%
%
\begin{abstract}
The state-space method is adapted to obtain three dimensional exact solutions for the static and damped dynamic behaviors of simply supported general laminates. The state-space method is written in a general form that permits to handle both cross-ply and antisymmetric angle-ply laminates. This general form also permits to obtain exact solutions for general laminates, albeit with some constraints. For the general case and for the static behavior, either an additive term is added to the load to simulate simply supported boundary conditions, or the plate bends in a particular way. For the dynamic behavior, the general case leads to pairs of natural frequencies for each order, with associated mode shapes.
\par
Finite element simulations have been performed to validate most of the results presented in this study. As the boundary conditions needed for the general case are not so straightforward, a specific discussion has been added. It is shown that these boundary conditions also work for the two aforementioned laminate classes. 
\par
The damped harmonic response of a non symmetrical isotropic sandwich is studied for different frequencies around the fundamental frequency. The static and undamped dynamic behaviors of the [-15/15], [0/30/0] and [-10/0/40] laminates are studied for various length-to-thickness ratios.
\end{abstract}
\end{frontmatter}
\input{01_Introduction.tex}
\input{02_Model.tex}
\input{03_FiniteElementSimulations.tex}
\input{04_Results.tex}
\input{05_Conclusion.tex}
\bibliographystyle{unsrt}
\bibliography{Ex3DSolLam}
\appendix
%
\input{06_Appendix.tex}
%
\label{dernierepage}
\end{document}

%% file: 01_Introduction.tex
\section{Introduction}
In the field of the study of multilayered anisotropic plates, exact 3D solutions occupy a particular place. They have been obtained for a consequent variety of problems including static mechanical behavior, undamped and damped dynamic behavior, sometimes coupled with initial stresses, thermal and piezoelectric loads. Solutions for simply supported plates are the easiest to obtain if the load is supposed to vary like a bi-sine function, because only a single term has to be handled for each quantity. For more general loadings, the decomposition of all quantities with respect to a basis must be performed, leading to systems of equations of growing size. That is the reason why these exact 3D solutions are of poor practical interest: easy to compute for unrealistic problems and hard to compute for practical problems. However, they are of particular interest from the theoretical point of view. Pagano's solution~\cite{Pagano1970}, Srinivas' solutions~\cite{Srinivas1970,Srinivas1970b} have been cited hundred of times in works of various nature like plate theories, numerical simulations, and experimental studies. These exact solutions had permitted to confirm or infirm numerous hypothesis, to enhance plate models, to validate finite element behavior, etc.
\par
In addition to these first studies, we can cite works for cylindrical shells~\cite{Kapuria2009,Wang2010} including interfacial damage~\cite{Chen2004}, works for thermal loading~\cite{Savoia1995}, works dealing with thermoelectroelastic coupling~\cite{Xu1995} which has been extended for the vibrations of initially stressed plates~\cite{Xu1997} and circular, annular, and sectorial plates~\cite{Xu2008}. Additional references can be found in~\cite{Chao2000}, and more recent works~\cite{Baillargeon2005,Kulikov2012}. The state-space method can also be used with functionally graded materials~\cite{Wu2012}, possibly including viscoelastic foundation~\cite{Hasheminejad2012}.
\par
For plates studies, solutions have first been obtained for cross-ply laminates in works like~\cite{Pagano1970,Srinivas1970,Srinivas1970b}. Since, many other studies have been concerned with this class of laminates. Exact solutions have also been obtained for antisymmetric angle ply laminates, first in reference~\cite{Noor1990}, and then in~\cite{Savoia1992, Savoia1995,Kulikov2012}. Among these works, references~\cite{Savoia1995,Kulikov2012} deal with both classes of laminates, but they are not treated exactly in the same manner: the basis of functions and the boundary conditions are chosen to fit the behavior of the considered class. In~\cite{Savoia1992}, authors present a method, based on an iterative variational approach, which can treat general lamination schemes. For general lamination schemes, only an approximate solution can be obtained, but this method provides the exact solution for cross-ply and antisymmetric angle ply laminates. In a more recent work~\cite{Kulikov2013}, authors also treat both class of laminates in the framework of electroelasticity with the help of a sampling surfaces method. The solution is dependent on the number of sampling surfaces that are considered. They extend the study to a laminate which is called \emph{unsymmetric angle-ply} in their paper, which is in fact --from the mechanical point of view-- an antisymmetrical angle-ply laminate bonded on the top and bottom surfaces by an isotropic layer. For this particular case, the exact solution can be obtained for the simply supported boundary condition.
\par
In this paper, the state-space method is used to obtain exact solutions for the bending of laminates. It is shown that the cross-ply and the antisymmetric angle-ply cases can be treated with the same process, including the same choice of basis functions and boundary conditions. At the end of the process, the two above cases naturally separates because null or particular constants values appear that makes the behavior of each type of laminate appropriate. Further, the same process can be applied to a general lamination sequence. It can give an exact solution for these laminates according to the following limitation: either the simply supported condition is no longer verified, or an additional term must be added to the loading.
\par
Three dimensional finite element simulations have been added to this study. A section is devoted to the study of the particular boundary conditions that must be applied to a 3D mesh to fit the corresponding boundary conditions of the problem. An additional interest of these simulations is to validate the above solving process, especially for damped structures and for general laminates for which no guarantee were given by previous works.
\par
Without loss of generality, the study is intentionally limited to cases involving a single order decomposition of all quantities.

%% file: 02_Model.tex
\section{Governing equations and solving method}\label{sec:Solving}
The solving procedure presented in this section belongs to the state-space method family. In the following, Greek subscripts take values $1$ or $2$ and Latin subscripts take values $1$, $2$ or $3$. Einstein's summation convention is used for subscripts only. The comma used as a subscript index means the partial derivative with respect to the directions corresponding to the following indexes. Displacements are supposed to be small enough to discard the nonlinear part of the strain tensor. 
\par
The plate is located in $(x,y)\in[0,a]\times[0,b]$. The laminate is composed of $N$ layers located between~$-h/2$ and $h/2$, where $h$ is the total height. The reference plane is taken as the $z=0$ plane. The $\ell$th layer is located between elevations $\zeta^{\ell-1}$ and $\zeta^{\ell}$, hence $\zeta^0=-h/2$ and $\zeta^N=h/2$.
\par
Each ply is supposed to have a linear orthotropic (visco)elastic behavior with $(x,y)$ as a plane of symmetry. This behavior is taken into account with the help of a complex fourth--order Hooke's tensor $C^{\ell}_{ijkl}$. In addition, all the fields are considered to have complex values. Volumetric forces are neglected. Hence, in the $\ell$--th layer:
\begin{equation}
  \label{eq:equilibrium}
  \left\{
    \begin{aligned}
      \varepsilon_{ij}&=\tfrac{1}{2}(u_{i,j}+u_{j,i})\\ 
      \sigma_{ij,j}&=\rho^{\ell} \ddot{u}_i 
    \end{aligned} 
  \right.
  \quad
  \text{with}
  \quad
  \sigma_{ij}=C^{\ell}_{ijkl}\varepsilon_{ij}\\ 
\end{equation}
At this time, the state-space method main idea is used. The system is reorganized, separating the derivatives with respect to $z$ from derivatives with respect to $x$ and~$y$:
\begin{equation}
  \label{eq:equilibrium2}
  \left\{
    \begin{aligned}
      u_{1,3}&=2\varepsilon_{13}-u_{3,1}\\ 
      u_{2,3}&=2\varepsilon_{23}-u_{3,2}\\ 
      u_{3,3}&=\varepsilon_{33}\\ 
      \sigma_{13,3}&=\rho^{\ell} \ddot{u}_1-\sigma_{11,1}-\sigma_{12,2}\\ 
      \sigma_{23,3}&=\rho^{\ell} \ddot{u}_2-\sigma_{21,1}-\sigma_{22,2}\\ 
      \sigma_{33,3}&=\rho^{\ell} \ddot{u}_3-\sigma_{31,1}-\sigma_{32,2} 
    \end{aligned} 
  \right.
\end{equation}
In order to eliminate $\varepsilon_{\alpha3}$, $\varepsilon_{33}$ and $\sigma_{\alpha\beta}$ from the above equations, Hooke's law is used:
\begin{equation}
\label{eq:GenPlaneStressHooke}
  \left\{
    \begin{aligned}
      \varepsilon_{\alpha3} &= 2S^{\ell}_{\alpha3\beta3}\sigma_{\beta3}\\\noalign{\medskip}
      \varepsilon_{33} &= \frac{1}{C^{\ell}_{3333}}(\sigma_{33}-C^{\ell}_{33\alpha\beta}\varepsilon_{\alpha\beta})\\ 
      \sigma_{\alpha\beta} &= C^{\ell}_{\alpha\beta\gamma\delta}\varepsilon_{\gamma\delta}+C^{\ell}_{\alpha\beta33}\varepsilon_{33} 
                            = Q^{\ell}_{\alpha\beta\gamma\delta}\varepsilon_{\gamma\delta}+\frac{C^{\ell}_{\alpha\beta33}}{C^{\ell}_{3333}}\sigma_{33}
    \end{aligned} 
  \right.
\end{equation}
where $S^{\ell}_{\alpha3\beta3}$ are components of the compliance tensor and $Q^{\ell}_{\alpha\beta\gamma\delta}=C^{\ell}_{\alpha\beta\gamma\delta}-C^{\ell}_{\alpha\beta33}C^{\ell}_{33\gamma\delta}/C^{\ell}_{3333}$ are the generalized plane stress stiffnesses\footnote{Note that the generalized plane stress hypothesis is not taken into account in this work, as it is generally the case for 2D plate models, but these well known quantities naturally appear here.}. Replacing $\varepsilon_{\alpha3}$, $\varepsilon_{33}$ and $\sigma_{\alpha\beta}$ into~\eqref{eq:equilibrium2} by means of formulas~\eqref{eq:GenPlaneStressHooke}, and then, replacing the strains $\varepsilon_{\alpha\beta}$ with the corresponding displacement derivatives lead to a system which only depends on $u_1$, $u_2$, $u_3$, $\sigma_{13}$, $\sigma_{23}$, $\sigma_{33}$ and their spatial and temporal derivatives.
\par
The method takes advantage of the decomposition of each quantity into a sum of a dyadic product of trigonometric functions of $x$ and $y$ like, taking $u_1$ as an example:
\begin{align}
  \label{eq:decomp}\nonumber
      u_1(x,y,z,t)=\sum_{m=1}^M \sum_{n=1}^N \left[u^{s,m,n}(z)\cos\left(\frac{m\pi}{a} x\right)\sin\left(\frac{n\pi}{b} y\right)\right.\\
        \left.+u^{a,m,n}(z)\sin\left(\frac{m\pi}{a} x\right)\cos\left(\frac{n\pi}{b} y\right)\right] e^{\text{j}\omega t}
\end{align}

where $u^{s,m,n}(z)$ and $u^{a,m,n}(z)$ are the $2\times M\times N$ components of $u(x,y,z,t)$ with respect to the trigonometric basis, which are in fact functions of $z$ that must be determined. For clarity, let us focus on a single term, corresponding to a choice of $m$ and $n$ in the above sum and let us introduce $\xi=m\pi/a$ and $\eta=n\pi/b$. From this point, superscripts $m,n$ and the time contribution are omitted for clarity. All quantities are expressed with the help of a trigonometric basis as follow:
\begin{equation}
\label{eq:functions}
  \left\{
    \begin{aligned}
      u_1(x,y,z)&=u^s(z)\cos(\xi x)\sin(\eta y)+u^a(z)\sin(\xi x)\cos(\eta y)\\
      u_2(x,y,z)&=v^s(z)\sin(\xi x)\cos(\eta y)+v^a(z)\cos(\xi x)\sin(\eta y)\\
      u_3(x,y,z)&=w^s(z)\sin(\xi x)\sin(\eta y)+w^a(z)\cos(\xi x)\cos(\eta y)\\
      \sigma_{13}(x,y,z)&=\sigma^s_{13}(z)\cos(\xi x)\sin(\eta y)+\sigma^a_{13}(z)\sin(\xi x)\cos(\eta y)\\
      \sigma_{23}(x,y,z)&=\sigma^s_{23}(z)\sin(\xi x)\cos(\eta y)+\sigma^a_{23}(z)\cos(\xi x)\sin(\eta y)\\
      \sigma_{33}(x,y,z)&=\sigma^s_{33}(z)\sin(\xi x)\sin(\eta y)+\sigma^a_{33}(z)\cos(\xi x)\cos(\eta y)
    \end{aligned} 
  \right.
\end{equation}
Introduction of these functions into the system previously built lead to a $12\times 12$ first--order differential system of equations of the form:
\begin{equation}
\label{eq:system}
  \frac{\partial}{\partial z}\left\{X(z)\right\}=\left[A^{\ell}\right]\left\{X(z)\right\}
\end{equation}
where:
\begin{align}
\label{eq:vector}\nonumber
  \left\{X(z)\right\}^T=\left\{u^s(z),v^s(z),w^s(z),\sigma^s_{13}(z),\sigma^s_{23}(z),\sigma^s_{33}(z),\right.\\
                         \left.u^a(z),v^a(z),w^a(z),\sigma^a_{13}(z),\sigma^a_{23}(z),\sigma^a_{33}(z)\right\}
\end{align}
and the matrix $\left[A^{\ell}\right]$ is given in~\ref{app:app1}. This system admits the solution:
\begin{equation}
\label{eq:solution}
  \left\{X(z)\right\}=e^{\left[A^{\ell}\right]z}\left\{C^{\ell}\right\}
\end{equation}
where $\left\{C^{\ell}\right\}$ is a vector which contains $12$ constants that must be determined with the boundary conditions. Depending on the position of the layer, these boundary conditions are either continuity conditions at the interfaces, either boundary conditions on the lower or on the upper face of the plate. Hence, at each of the $N-1$ interfaces $z=\zeta^{\ell}\text{, }\ell\in<1,N-1>$, the continuity of all the functions of~\eqref{eq:vector} gives $12$ independent conditions, which can be summarized as:
\begin{equation}
\label{eq:interface}
  \left\{X(\zeta^{\ell}_-)\right\}=\left\{X(\zeta^{\ell}_+)\right\}\text{, }\ell\in<1,N-1>
\end{equation}
or:
\begin{equation}
\label{eq:interface2}
  e^{\left[A^{\ell}\right]\zeta^{\ell}}\left\{C^{\ell}\right\}=e^{\left[A^{\ell+1}\right]\zeta^{\ell}}\left\{C^{\ell+1}\right\}\text{, }\ell\in<1,N-1>
\end{equation}
On the lower and upper faces, the three displacements remain unknown. Let the three stresses of~\eqref{eq:functions} have the following prescribed values: 
\begin{equation}
\label{eq:functions}
  \left\{
    \begin{aligned}
      \sigma_{13}(x,y,-\tfrac{h}{2})&=\sigma_{13}(x,y,+\tfrac{h}{2})=0\\
      \sigma_{23}(x,y,-\tfrac{h}{2})&=\sigma_{23}(x,y,+\tfrac{h}{2})=0\\
      \sigma_{33}(x,y,-\tfrac{h}{2})&=+\tfrac{q}{2}\sin(\xi x)\sin(\eta y)+\tfrac{p}{2}\cos(\xi x)\cos(\eta y)\\
      \sigma_{33}(x,y,+\tfrac{h}{2})&=-\tfrac{q}{2}\sin(\xi x)\sin(\eta y)-\tfrac{p}{2}\cos(\xi x)\cos(\eta y)
    \end{aligned} 
  \right.
\end{equation}
This leads to $6 \times 2 = 12$ additional conditions:
\begin{equation}
\label{eq:lowupp}
  \left\{
    \begin{alignedat}{3}
      \left\{X(-\tfrac{h}{2})\right\}^T&=&\left\{u^s(-\tfrac{h}{2}),v^s(-\tfrac{h}{2}),w^s(-\tfrac{h}{2}),0,0,+\tfrac{q}{2},\right. \\
                                       & &\left. u^a(-\tfrac{h}{2}),v^a(-\tfrac{h}{2}),w^a(-\tfrac{h}{2}),0,0,+\tfrac{p}{2}\right\} \\
      \left\{X(+\tfrac{h}{2})\right\}^T&=&\left\{u^s(+\tfrac{h}{2}),v^s(+\tfrac{h}{2}),w^s(+\tfrac{h}{2}),0,0,-\tfrac{q}{2},\right. \\
                                       & &\left. u^a(+\tfrac{h}{2}),v^a(+\tfrac{h}{2}),w^a(+\tfrac{h}{2}),0,0,-\tfrac{p}{2}\right\}
    \end{alignedat}
  \right.
\end{equation}
The term $p$ is null, except for general laminates as we shall see later. Hence, there are $12 \times N$ independent equations in~\eqref{eq:interface2} and~\eqref{eq:lowupp} for the $12 \times N$ unknowns:
\begin{equation}
\label{eq:unknows}
  \left\{
    \begin{aligned}
      &u^s(-\tfrac{h}{2}),v^s(-\tfrac{h}{2}),w^s(-\tfrac{h}{2}),u^a(-\tfrac{h}{2}),v^a(-\tfrac{h}{2}),w^a(-\tfrac{h}{2}), \\
      &C_1^{\ell},C_2^{\ell},\dots,C_{12}^{\ell}\text{, with }\ell\in<1,N-1>, \\
      &u^s(+\tfrac{h}{2}),v^s(+\tfrac{h}{2}),w^s(+\tfrac{h}{2}),u^a(+\tfrac{h}{2}),v^a(+\tfrac{h}{2}),w^a(+\tfrac{h}{2})
    \end{aligned}
  \right.
\end{equation}
Thus, this linear system can be solved, and the solution is computed.
\subsection{The case of general laminates}\label{sec:CaseGenLam}
A detailed discussion for the specific simply supported boundary conditions, covering all the cases, is given in section~\ref{sec:FEs}. We will only focus in this section to the particular case of general laminates. Applying the above process lead, for the general case, to a deflection which does not verify the simply supported conditions. The $w^a(z)$ function does not verify $w^a(0)=0$, which means that the corners of the plate moves along the $z$ direction. The solution remains exact because no other terms are needed to balance the equations, even for the more general laminate. This boundary condition will be designated in the following by \emph{globally simply supported} (GSS).
\par
This particular set of boundary conditions is kept, in our study, for the search of natural frequencies. We shall see that each order $(m,n)$ is associated with two distinct natural frequencies. The mode shapes associated with these two frequencies are like cylindrical bending in the $\pm45°$ directions. These laminates have different bending stiffnesses in the $\pm45°$ directions, which explains the splitting of frequencies.
\par
However, for the static case, it is possible to simulate an equivalent of the simply supported condition if a bi-cosine term with amplitude $p$ is added to the loading. The above system is changed in the following way: $p$, which is prescribed to 0 in the classical cases is now an unknown, and the equation $C^{\ell_{\text{ref}}}_9=0$ is added to equations~\eqref{eq:interface2} and~\eqref{eq:lowupp}. This added equation forces $w^a(0)$ to be null. Indeed, for the layer(s) $\ell_{\text{ref}}$ which contains the reference plane, setting $z=0$ in equation~\eqref{eq:solution} shows that $w^a(0)=C^{\ell_{\text{ref}}}_9$. This solution will be designated in the following by \emph{simulated simply supported} (SSS). For dynamic purposes, it is possible to search for a value of $p$ for each frequency of a response curve, but the meaning of such results is quite poor. 
\subsection{Dynamic behavior}
We shall restrain here the study to harmonic solicitations, although state methods have been used for general time response~\cite{Wei2006}. For harmonic solicitations, one can search for the response to a load at a given frequency or search for natural frequencies. The response to a given frequency to a bi-sine load is computed using the same procedure as the static case, the only difference is the non null value for $\omega$ in matrices $\left[A^{\ell}\right]$. The search for natural frequencies is a more difficult task. Several works~\cite{Xu1997,Qing2008} have described an exact procedure, for undamped structures, which consists in searching for roots of the determinant of a matrix. This procedure has been successfully used for matrices of size $6 \times 6$, but we have been unable to make it work for our $12 \times 12$ matrices. Let us summarize the method: it starts with the transfer matrix which links the upper face to the bottom face according to:
\begin{equation}
  \label{eq:TransferMatrix}
  \left\{X(\tfrac{h}{2})\right\}=\left[\Lambda(\omega)\right] \left\{X(-\tfrac{h}{2})\right\} = \prod_{\ell=N}^{1} e^{\left[A^{\ell}\right]h^{\ell}}\left\{X(-\tfrac{h}{2})\right\}
\end{equation}
A natural frequency for an undamped plate corresponds to non null displacements $u$, $v$, and $w$ for a null load $q$. Hence the sub-matrix formed with the lines (1,2,3,7,8,9) and the columns (4,5,6,10,11,12) of the matrix $\left[\Lambda(\omega)\right]$ must have a null determinant.
\par
This equation is transcendental with respect to $\omega$ and leads to an infinite number of natural frequencies for each couple $(m,n)$. In some works, authors~\cite{Srinivas1970b} have noticed that each of these natural frequencies corresponds with a specific transverse mode, and the lower of them corresponds to the bending natural frequency of the mode $(m,n)$.
\par
All the data in $\left[A^{\ell}\right]$ can be replaced with their corresponding values except the angular frequency $\omega$ which remains the unknown. Unfortunately, because of the size of $12\times12$ of the matrices $\left[A^{\ell}\right]$ in our case, we have been unable to compute analytically their exponential, and so unable to obtain the matrix $\left[\Lambda(\omega)\right]$.
\par
Hence, the natural frequencies are searched with an iterative procedure. It can be noticed that other authors~\cite{Kapuria2005} have also proposed to search natural frequencies by means of an iterative process. A bi-sine load of order $(m,n)$ is applied and the complex input power is computed. This complex power is $P_{inc}=\tfrac{abq}{8}(w^s(\tfrac{h}{2})-w^s(\tfrac{-h}{2}))\text{j}$. It has been shown, for example in reference~\cite{Castel2012}, that the imaginary part of this input power is related to the Lagrangian of the plate, when the real part corresponds to the dissipated power. Hence, the frequency is varied in order to find the one which gives a null value of the Lagrangian, which corresponds to the searched mode(s). The process is:
\begin{itemize}
  \item for a given $(m,n)$, a (low) starting frequency $\omega_0$ is set, the system is solved for a bi-sine load of order $(m,n)$, and the Lagrangian is computed;
  \item at each following iteration, frequency is multiplied by $k$ so $\omega_{i+1}=k\omega_{i}$, and the corresponding Lagrangian is computed until a change on the sign is detected, say for $i=\iota$;
  \item a final step consists in a dichotomic search between $\omega_{\iota}$ and $\omega_{\iota-1}$ to find the natural frequency, the process being stopped for a required precision. \end{itemize}
This process is easy to implement, and gives excellent results. The value for $k$ can be set to $1.8$ for laminates but attention should be paid:
\begin{itemize}
  \item for sandwiches, especially if the core have very low mechanical characteristics compared to the skins. In these cases, different transverse modes may have close frequencies, hence the value for $k$ must be diminished to avoid the ``jump'' of two frequencies,
  \item for general laminates, because, as told in section~\ref{sec:CaseGenLam}, two modes exist for each value of $(m,n)$. A straightforward modification of the process previously described can help to manage these cases.
\end{itemize}

%% file: 03_FiniteElementSimulations.tex
\section{Finite element simulations}\label{sec:FEs}
Three dimensional finite element simulations have been performed for most of the examples presented in section~\ref{sec:Results}. This was initially done in order to verify the solutions of the analytical method, and these results were not scheduled to figure in the paper. However, the task was not as easier than expected, especially the choice of boundary conditions. In addition, results for general laminates cannot be compared with data of the literature. It is the reason why these results have been finally given, and the specific 3D finite element boundary conditions are detailed in this section.
\par
Let us start with the general form of the displacement functions~:
\begin{equation}
\label{eq:functions2}
  \left\{
    \begin{array}{l}
      u(x,y,z)=u^s(z)\cos(\xi x)\sin(\eta y)+u^a(z)\sin(\xi x)\cos(\eta y)\\
      v(x,y,z)=v^s(z)\sin(\xi x)\cos(\eta y)+v^a(z)\cos(\xi x)\sin(\eta y)\\
      w(x,y,z)=w^s(z)\sin(\xi x)\sin(\eta y)+w^a(z)\cos(\xi x)\cos(\eta y)
    \end{array} 
  \right.
\end{equation}
This general form works for all laminates, but three cases should be distinguished: the cross-ply case, the antisymmetric angle-ply case and the general case. 
\subsection{The cross-ply case} 
For this class of problems, the $u^a(z)$, $v^a(z)$, and $w^a(z)$ functions are found to be null functions when the present method is applied. All the previous studies which have treated cross-play laminates have ignored the corresponding terms in formula~\eqref{eq:functions2}. Let us examine what happens at boundaries $x=0,a$ and $y=0,b$:
\begin{equation}
\label{eq:bound_cross_ply}
  \left\{
    \begin{array}{l}
      u(x,0,z)=u(x,b,z)=0\\
      v(0,y,z)=v(a,y,z)=0\\
      w(x,0,z)=w(x,b,z)=w(0,y,z)=w(a,y,z)=0
    \end{array} 
  \right.
\end{equation}
These particular simply supported conditions are generally applied for this class of problem, they permit to a laminate which exhibits a membrane-bending coupling to bend with no membrane constraint. Further, they are easy to apply to a finite element simulation. They are also compatible with the study of a quarter of the plate, with the help of appropriate symmetry boundary conditions applied to the cut edges.
\subsection{The antisymmetric angle-ply case}\label{sec:antisymcase}
For this class of problems, the $u^a(z)$, $v^a(z)$, and $w^a(z)$ functions are no longer null functions, hence, at boundaries $x=0,a$, and $y=0,b$, we have now:
\begin{equation}
\label{eq:bound_anti_sym}
  \left\{
    \begin{array}{l}
      u(x,0,z)=u^a(z)\sin(\xi  x) \quad \text{and} \quad u(x,b,z)=-u^a(z)\sin(\xi  x)\\
      v(0,y,z)=v^a(z)\sin(\eta y) \quad \text{and} \quad v(a,y,z)=-v^a(z)\sin(\eta y)\\
      w(x,0,z)=w^a(z)\cos(\xi  x) \quad \text{and} \quad w(x,b,z)=-w^a(z)\cos(\xi  x)\\
      w(0,y,z)=w^a(z)\cos(\eta y) \quad \text{and} \quad w(a,y,z)=-w^a(z)\cos(\eta y)
    \end{array} 
  \right.
\end{equation}
In this form, these conditions are no longer useful for finite element simulation, but we can also write them:
\begin{equation}
\label{eq:bound_anti_sym2}
  \left\{
    \begin{array}{l}
      u(x,0,z)+u(x,b,z)=0\\
      v(0,y,z)+v(a,y,z)=0\\
      w(x,0,z)+w(x,b,z)=0\\
      w(0,y,z)+w(a,y,z)=0
    \end{array} 
  \right.
\end{equation}
This particular set of conditions is clearly more complex than the previous one. In fact it is also a more general one, because, as one could see, the previous set is included into this one. Further, this set could also be applied to cross-ply laminates and it will work perfectly. Fortunately, many finite element softwares, like Cast3M~\cite{Cast3M2012} used in this study, permit to apply such boundary conditions. However, they are not compatible with the study of a quarter of the plate. It is also interesting to notice that when applying this set to antisymmetric cross-ply laminates, the $w^a(z)$ function verifies $w^a(z)=-w^a(-z)$ which implies $w^a(0)=0$. That means that the plate remains supported on its edges, the $w^a(z)$ function only permit non null $\varepsilon_{33}(z)$ on the edges. We shall also notice that there is no need to prescribe the $w^a(0)=0$ condition to the finite element problem, the antisymmetry of the laminate is sufficient to give the good result. 
\subsection{The general case}\label{sec:genecase}
As told in section~\ref{sec:CaseGenLam}, the previous boundary conditions, which works for both cross-ply and antisymmetric angle-ply laminates, leads, for the general laminate, to the \emph{globally simply supported condition} (GSS). The plate is no longer strictly supported on its edges. To be more precise, it is supported in a particular way which corresponds to the last two relations of~\eqref{eq:bound_anti_sym2}, but the antisymmetry of the $w^a(z)$ function which was encountered in the previous case does not appear in the general case. That means that the plate could have a non null deflection on its edges, with respect of an antisymmetry with the opposite edge. Applying this set of boundary conditions to the finite element simulation of a general laminate will give the same solution than the analytical process.
\par
However, for the static case, another approach has been chosen in this study. As described in section~\ref{sec:CaseGenLam}, the simulated simply supported (SSS) condition is prescribed for a single order loading (of order (1,1) in our case) with the help of a supplementary $\cos(\xi x)\cos(\eta y)$ term added to the loading. The amplitude of this term, $p$, which appears in formula~\eqref{eq:lowupp}, must be adapted to each case (laminate, length-to-thickness ratio, order of the loading\dots). This is straightforward for the analytical solving procedure because $p$ becomes an unknown of the system as well as those of equation~\eqref{eq:unknows}, but for the finite element method, it necessitates two computations. The first is done with $p=0$ and the second with $p=p_0$ giving two different values for the deflection of a corner of the plate. Then the value of $p$ that gives a null value of the deflection of this corner is calculated and the deflection can be found accounting to the linearity (a third computation is not necessary).
\subsection{Mesh considerations}\label{sec:MeshCons}
For studies involving composite materials, finite element computations have been made with a regular $24\times24\times24$ 20-nodes hexaedron mesh, except for the computation of the transverse stresses in the case of a length-to-thickness ratio $a/h$ of 100. In this case, because elements have a high length-to-thickness ratio, the stresses are not properly evaluated. A local refinement of the mesh at the points where the stresses are computed, with elements four times smaller than those at the corners of the plate, have been necessary to obtain correct results that can been seen in figures~\ref{fig:m15p15various},~\ref{fig:p0p30p0various} and~\ref{fig:m10p0p40various}. Note that even for $a/h=100$, transverse displacements and natural frequencies presented in the tables have been computed with the regular mesh, and they are in very good agreement with the analytical solution. However, one can notice that for the higher value of $a/h$, deflections and natural frequencies are a little less good than for other values of $a/h$, this is also due to the bad shape of the elements for $a/h=100$.
\par
For the first study, the viscoelastic behavior is considered. The use of complex numbers leads to a system of double size. Hence, the number of elements along a side of the plate has been limited to $16$. However, due to symmetry considerations which does not apply for the other presented studies, it is possible here to consider only a quarter of the plate. That permits to have approximatively the same precision than for the other studies. The mesh has been made with 20-nodes hexaedrons and 15-nodes pentaedrons with a refinement at the location where the stresses are evaluated. The small elements are $4$ times smaller than the biggest. This plate has $3$ layers, the number of elements in the $z$ direction has been set to $18=12+3+3$.

%% file: 04_Results.tex
\section{Results}\label{sec:Results}
In this section, laminates made of different materials are studied. First, an isotropic unsymmetrical sandwich is considered. Although this laminate belongs to the cross-ply family, the reason why it is presented here is mainly to show how viscoelastic behavior for an harmonic solicitation is effectively handled by the present method. The sandwich is made of aluminum alloy and viscoelastic polymer which properties are:

\begin{itemize}
	\item aluminum alloy: $E=72.4~\text{GPa}$, $\nu=0.34$, $\rho=2780~\text{kg.m}^{-3}$, $\eta=0.005$. 
	\item viscoelastic material: $E=2.30~\text{MPa}$, $\nu=0.45$, $\rho=1015~\text{kg.m}^{-3}$, $\eta=1$.
\end{itemize}
where $\eta$ is the damping ratio.
\par
All the other studies involve laminates made of plies of equal thicknesses. Each ply is made up of transversely isotropic composite material which characteristics are:
\begin{itemize}
	\item composite material: $E_L = 25E_T$, $E_T = 10^6$, $\nu_{LT}=0.25$, $\nu_{TT}=0.25$, $G_{LT}=0.5E_T$, $G_{TT}=0.2E_T$, $\rho=1500$.
\end{itemize}

To cover the anti-symmetrical case, a study of a [-15/15] laminate is performed. Static and dynamic behavior are considered and transverse stresses are plotted for various length-to thickness ratios. Finally, the same study is performed for both [0/30/0] and [-10/0/40] laminates, showing the ability of the method for the most general case.
\par
Nondimensionalized values are obtained with formulas:
\begin{equation}
\label{eq:Nondimensionalized}
  \left\{
    \begin{array}{lll}
      \text{Displacements} & : &\overline{u}_i = 100 \frac{E_2 h^3}{(-q) a^2 b^2}u_i\\\noalign{\medskip}
      \text{Stresses}      & : &\overline{\sigma}_{ij} = 10 \frac{h}{q a}\sigma_{ij}\\\noalign{\medskip}
      \text{Frequencies}   & : &\overline{\omega}= ab \sqrt{\frac{\rho }{E_2 h^2}}\omega;
    \end{array} 
  \right.
\end{equation}

\subsection{Damped dynamic behavior of an unsymmetrical sandwich}
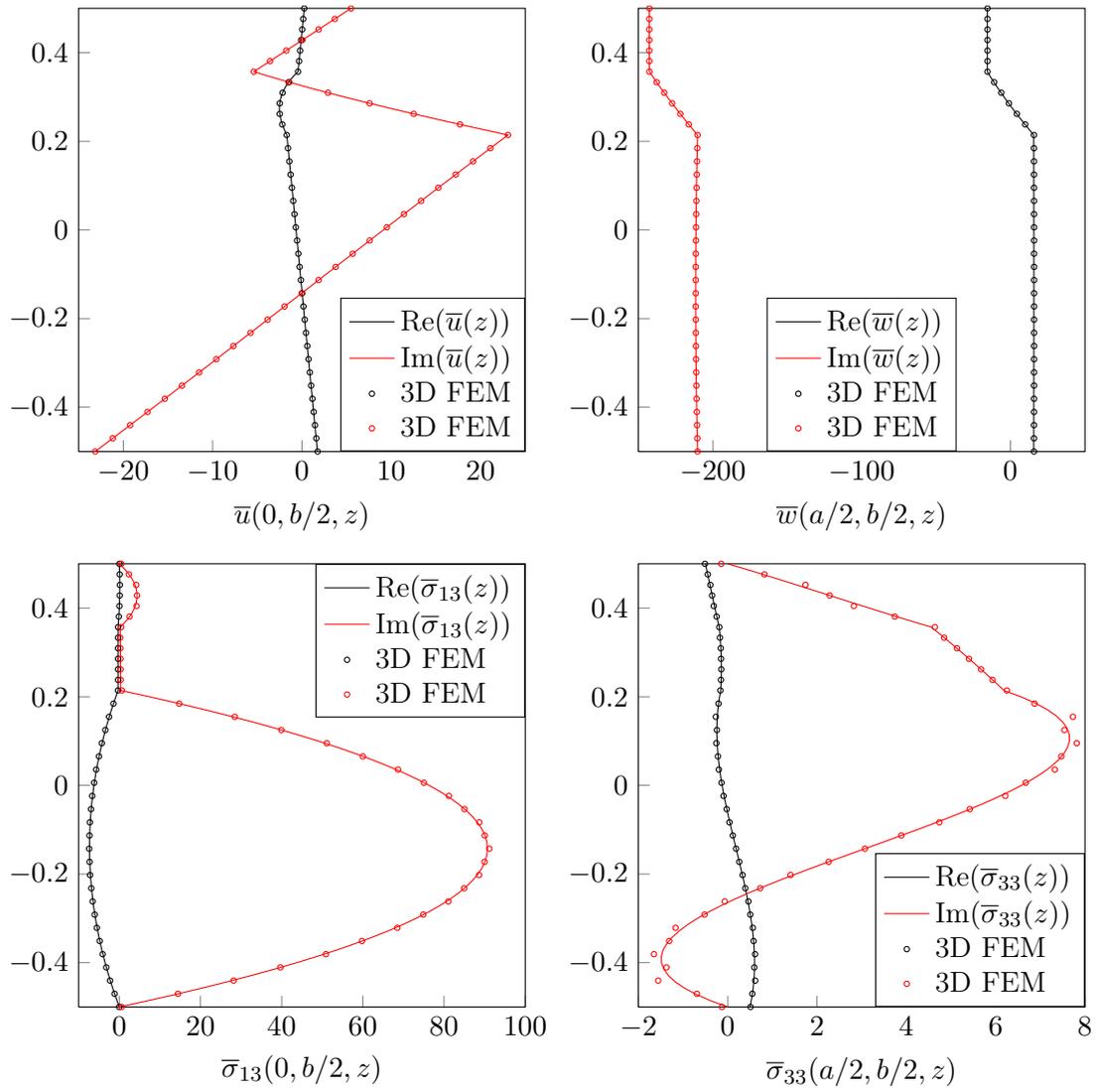
\begin{figure}
  \centering
  \iftoggle{submission}{
    \includegraphics{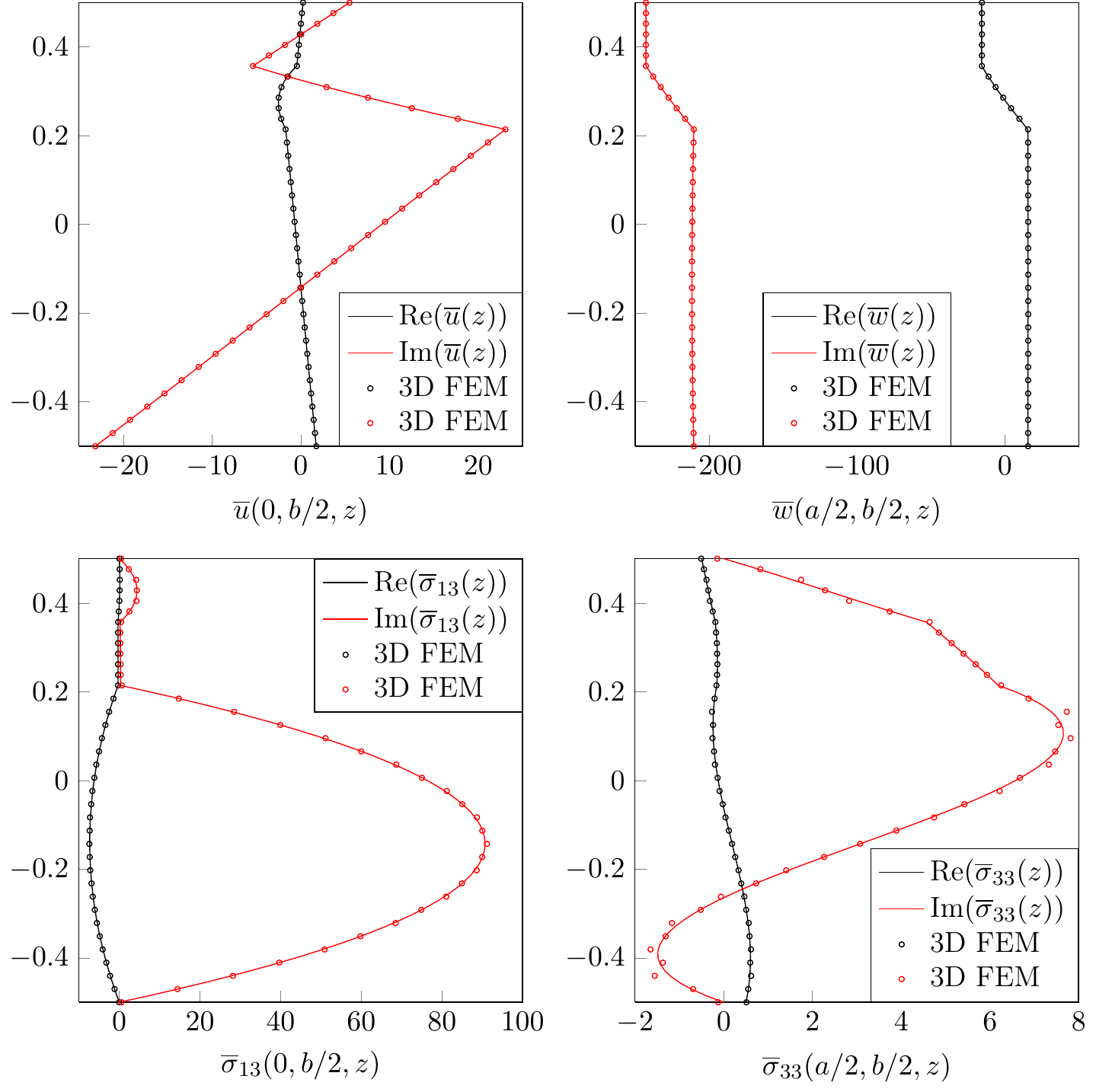}
  }{
    \begin{tikzpicture}
      \begin{groupplot}
        [
            group style=
                {
                group size =2 by 2,
                xlabels at=edge bottom,
                ylabels at=edge left,
                horizontal sep = 1.5 cm, 
                vertical sep = 1.5 cm 
                },
            width=7.5cm,height=7.5cm,
            tickpos=left,        
            cycle list={{solid, black},{solid, red},{black, only marks, mark=o},{red, only marks, mark=o}},
            transpose legend,
            legend cell align = left,
        ]
        \nextgroupplot[width=7.5cm,height=7.5cm,xlabel={$\overline{u}(0,b/2,z)$},
          xmin=-25, xmax=25,ymin=-0.5,ymax=0.5,
          every axis legend/.append style={at={(1.0,0.0)},anchor=south east}
        ]
        \addplot file {Images/ISD112_Re(us)_h=1.000_freq=30.3.txt};
        \addplot file {Images/ISD112_Im(us)_h=1.000_freq=30.3.txt};
        \addplot file {Images/ISD112_Castem_Re(us)_h=1.000_freq=30.3.txt};
        \addplot file {Images/ISD112_Castem_Im(us)_h=1.000_freq=30.3.txt};
        \legend{$\text{Re}(\overline{u}(z))$, $\text{Im}(\overline{u}(z))$, 3D FEM, 3D FEM}
        \nextgroupplot[width=7.5cm,height=7.5cm,xlabel={$\overline{w}(a/2,b/2,z)$},
          xmin=-250, xmax=50,ymin=-0.5,ymax=0.5,
          every axis legend/.append style={at={(0.5,0.0)},anchor=south}
        ]
        \addplot file {Images/ISD112_Re(ws)_h=1.000_freq=30.3.txt};
        \addplot file {Images/ISD112_Im(ws)_h=1.000_freq=30.3.txt};
        \addplot file {Images/ISD112_Castem_Re(ws)_h=1.000_freq=30.3.txt};
        \addplot file {Images/ISD112_Castem_Im(ws)_h=1.000_freq=30.3.txt};
        \legend{$\text{Re}(\overline{w}(z))$, $\text{Im}(\overline{w}(z))$, 3D FEM, 3D FEM}
        \nextgroupplot[width=7.5cm,height=7.5cm,xlabel={$\overline{\sigma}_{13}(0,b/2,z)$},
          xmin=-10, xmax=100,ymin=-0.5,ymax=0.5,
          every axis legend/.append style={at={(1,1)},anchor=north east }
        ]
        \addplot file {Images/ISD112_Re(S13s)_h=1.000_freq=30.3.txt};
        \addplot file {Images/ISD112_Im(S13s)_h=1.000_freq=30.3.txt};
        \addplot file {Images/ISD112_Castem_Re(S13s)_h=1.000_freq=30.3.txt};
        \addplot file {Images/ISD112_Castem_Im(S13s)_h=1.000_freq=30.3.txt};
        \legend{$\text{Re}(\overline{\sigma}_{13}(z))$, $\text{Im}(\overline{\sigma}_{13}(z))$, 3D FEM, 3D FEM}
        \nextgroupplot[width=7.5cm,height=7.5cm,xlabel={$\overline{\sigma}_{33}(a/2,b/2,z)$},
          xmin=-2, xmax=8,ymin=-0.5,ymax=0.5,
          every axis legend/.append style={at={(1,0)},anchor=south east }
        ]
        \addplot file {Images/ISD112_Re(S33s)_h=1.000_freq=30.3.txt};
        \addplot file {Images/ISD112_Im(S33s)_h=1.000_freq=30.3.txt};
        \addplot file {Images/ISD112_Castem_Re(S33s)_h=1.000_freq=30.3.txt};
        \addplot file {Images/ISD112_Castem_Im(S33s)_h=1.000_freq=30.3.txt};
        \legend{$\text{Re}(\overline{\sigma}_{33}(z))$, $\text{Im}(\overline{\sigma}_{33}(z))$, 3D FEM, 3D FEM}
      \end{groupplot}%
    \end{tikzpicture}
  }
    \caption{Variation of the dimensionless displacements $u$ and $w$ and the dimensionless stresses $\sigma_{13}$ and $\sigma_{33}$ through the thickness of the sandwich plate excited at the dimensionless fundamental frequency of $\overline{\omega}_0=3.73157$ by an unitary (1,1) bi-sine load equally distributed on the upper and bottom faces.}
	\label{fig:ISD112}
\end{figure}

\begin{figure}
  \centering
  \iftoggle{submission}{
    \includegraphics{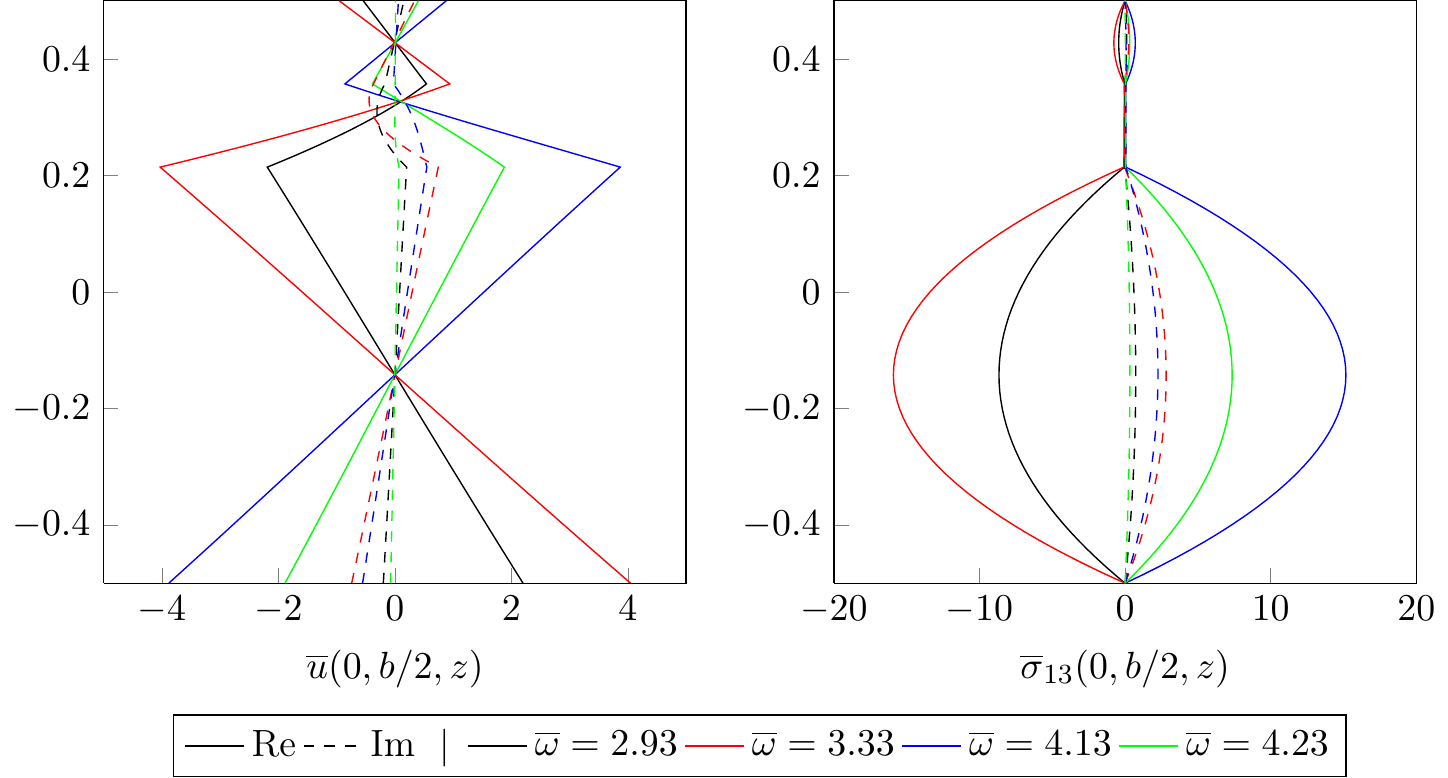}
  }{
    \begin{tikzpicture}
      \begin{groupplot}
        [
            group style=
                {
                group size =2 by 1,
                xlabels at=edge bottom,
                ylabels at=edge left,
                horizontal sep = 1.5 cm 
                },
            width=7.5cm,height=7.5cm,
            tickpos=left,        
            cycle list={{solid, black},{solid, red},{solid,blue},{solid,green},{dashed, black},{dashed, red},{dashed,blue},{dashed,green}},
            legend columns=-1,
            transpose legend,
            legend cell align = left,
        ]
        \nextgroupplot[xlabel={$\overline{u}(0,b/2,z)$},
          xmin=-5, xmax=5,ymin=-0.5,ymax=0.5,
          legend to name = legend1_ISD112
        ]
        \addlegendimage{solid, black};\addlegendentry{Re};
        \addlegendimage{dashed, black};\addlegendentry{Im};
        \addlegendimage{empty legend};\addlegendentry{$\;|\;$};
        \addplot file {Images/ISD112_Re(Us)_h=1.0_pulsadim=2.93.txt};\addlegendentry{$\overline{\omega}=2.93$};
        \addplot file {Images/ISD112_Re(Us)_h=1.0_pulsadim=3.33.txt};\addlegendentry{$\overline{\omega}=3.33$};
        \addplot file {Images/ISD112_Re(Us)_h=1.0_pulsadim=4.13.txt};\addlegendentry{$\overline{\omega}=4.13$};
        \addplot file {Images/ISD112_Re(Us)_h=1.0_pulsadim=4.53.txt};\addlegendentry{$\overline{\omega}=4.23$};
        \addplot file {Images/ISD112_Im(Us)_h=1.0_pulsadim=2.93.txt};
        \addplot file {Images/ISD112_Im(Us)_h=1.0_pulsadim=3.33.txt};
        \addplot file {Images/ISD112_Im(Us)_h=1.0_pulsadim=4.13.txt};
        \addplot file {Images/ISD112_Im(Us)_h=1.0_pulsadim=4.53.txt};
        \nextgroupplot[xlabel={$\overline{\sigma}_{13}(0,b/2,z)$},
          xmin=-20, xmax=20,ymin=-0.5,ymax=0.5
        ]
        \addplot file {Images/ISD112_Re(S13s)_h=1.0_pulsadim=2.93.txt};
        \addplot file {Images/ISD112_Re(S13s)_h=1.0_pulsadim=3.33.txt};
        \addplot file {Images/ISD112_Re(S13s)_h=1.0_pulsadim=4.13.txt};
        \addplot file {Images/ISD112_Re(S13s)_h=1.0_pulsadim=4.53.txt};
        \addplot file {Images/ISD112_Im(S13s)_h=1.0_pulsadim=2.93.txt};
        \addplot file {Images/ISD112_Im(S13s)_h=1.0_pulsadim=3.33.txt};
        \addplot file {Images/ISD112_Im(S13s)_h=1.0_pulsadim=4.13.txt};
        \addplot file {Images/ISD112_Im(S13s)_h=1.0_pulsadim=4.53.txt};
      \end{groupplot}%
      \node at ($(group c1r1.south)!0.5!(group c2r1.south)+(0,0cm)$) [inner sep=0pt,anchor=north, yshift=-8ex] {\pgfplotslegendfromname{legend1_ISD112}};
    \end{tikzpicture}
  }
  \caption{Variation of the dimensionless displacements $u$ and the dimensionless stresses $\sigma_{13}$ through the thickness of the sandwich plate at various frequencies around the dimensionless fundamental frequency of $\overline{\omega}_0=3.73157$.}
  \label{fig:ISD112_freq}
\end{figure}

The sandwich plate is made of three layers which materials are aluminum alloy for the skins and viscoelastic material for the core. The plate is a square with sides of length $a=b$, and with length-to-thickness ratio $a/h=10$. The layers have corresponding thicknesses $5h/7$, $h/7$, and $h/7$. The first natural bending frequency has been found to be $\overline{\omega}_0=3.73157$. Figure~\ref{fig:ISD112} shows the variation through the thickness of the nondimensionalized displacements $u$ and $w$ and stresses $\sigma_{13}$ and $\sigma_{33}$ when the plate is excited by a (1,1) bi-sine load at the fundamental frequency. Note that this does not correspond to a modal shape because the method used in this paper to find natural frequencies is based on the search of null values for the Lagrangian for an given load. This is the reason why values for $\sigma_{33}(\pm h/2)$ are not null. Figure~\ref{fig:ISD112_freq} shows the variation through the thickness of the nondimensionalized displacements $u$ and the stresses $\sigma_{13}$ when the plate is excited by a (1,1) bi-sine load at various frequencies around the fundamental frequency.
\iftoggle{submission}{}{\tikzsetnextfilename{ISD112}}
\iftoggle{submission}{}{\tikzsetnextfilename{ISD112_freq}}
\subsection{Static and dynamic behavior of a [-15/15] laminate}
\begin{figure}
	\centering
  \iftoggle{submission}{
    \includegraphics{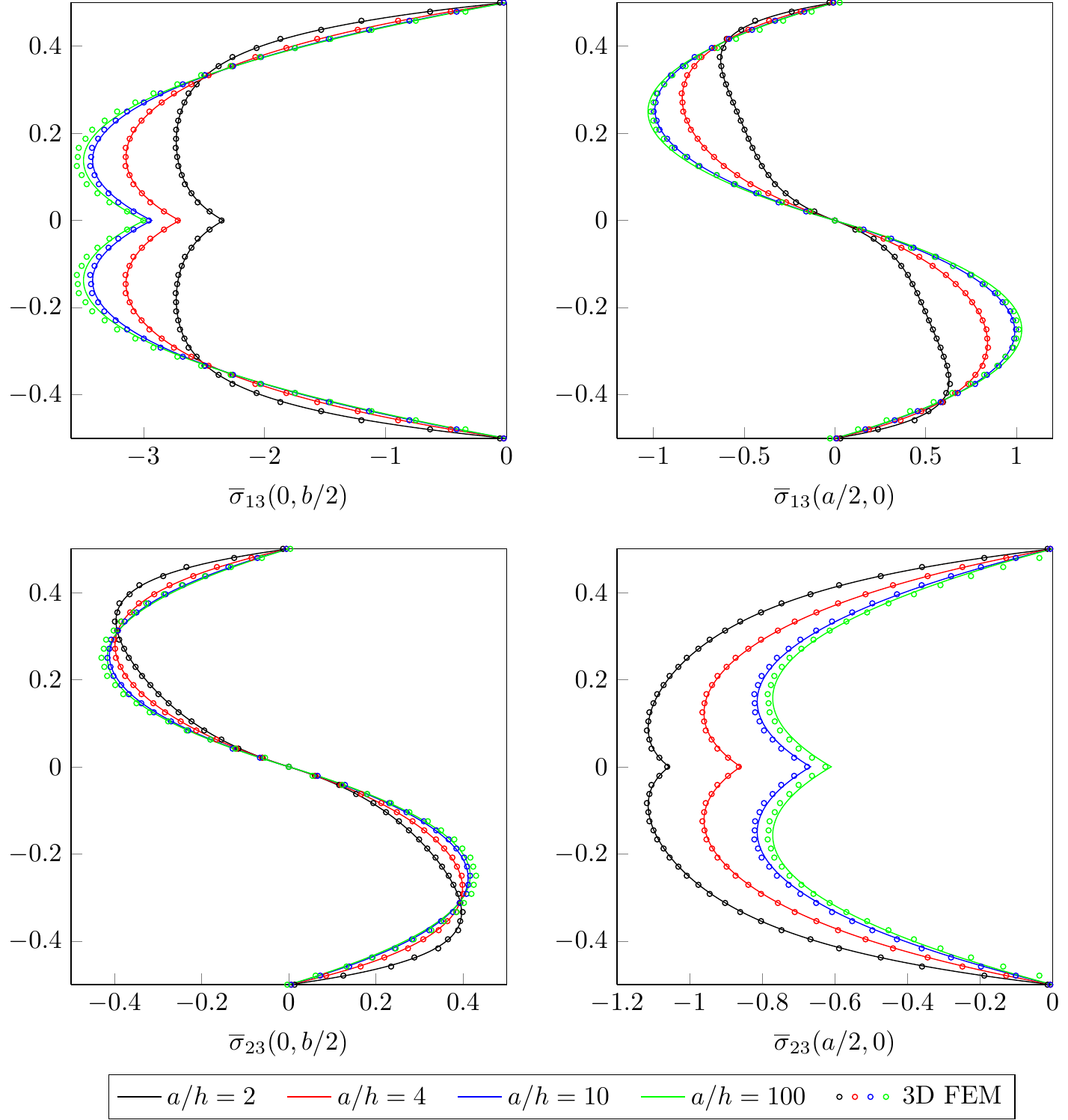}
  }{
    \begin{tikzpicture}
      \begin{groupplot}
        [
            group style=
                {
                group size =2 by 2,
                xlabels at=edge bottom,
                ylabels at=edge left,
                horizontal sep = 1.5 cm,
                vertical sep = 1.5 cm
                },
            width=7.5cm,height=7.5cm,
            tickpos=left,        
            cycle list={{solid, black},{solid, red},{solid, blue},{solid, green},{black, only marks, mark=o},
                        {red, only marks, mark=o},{blue, only marks, mark=o},{green, only marks, mark=o}},
            legend columns=-1,
            transpose legend,
            legend cell align = left,
        ]
        \nextgroupplot[width=7.5cm,height=7.5cm,xlabel={$\overline{\sigma}_{13}(0,b/2)$},
          xmin=-3.6, xmax=0,ymin=-0.5,ymax=0.5,
          legend to name = legend1_m15p15
        ]
        \addplot file {Images/m15p15_Kulikov_S13s2.txt};
        \addplot file {Images/m15p15_Kulikov_S13s4.txt};
        \addplot file {Images/m15p15_Kulikov_S13s10.txt};
        \addplot file {Images/m15p15_Kulikov_S13s100.txt};
        \addplot file {Images/m15p15_Kulikov_Castem_S13s2.txt};
        \addplot file {Images/m15p15_Kulikov_Castem_S13s4.txt};
        \addplot file {Images/m15p15_Kulikov_Castem_S13s10.txt};
        \addplot file {Images/m15p15_Kulikov_Castem_S13s100.txt};
        \legend{$a/h=2\quad$, $a/h=4\quad$, $a/h=10\quad$, $a/h=100\quad$, $$, $$, $$, ~3D FEM}
        \nextgroupplot[width=7.5cm,height=7.5cm,xlabel={$\overline{\sigma}_{13}(a/2,0)$},
          xmin=-1.2, xmax=1.2, ymin=-0.5, ymax=0.5,
        ]
        \addplot file {Images/m15p15_Kulikov_S13a2.txt};
        \addplot file {Images/m15p15_Kulikov_S13a4.txt};
        \addplot file {Images/m15p15_Kulikov_S13a10.txt};
        \addplot file {Images/m15p15_Kulikov_S13a100.txt};
        \addplot file {Images/m15p15_Kulikov_Castem_S13a2.txt};
        \addplot file {Images/m15p15_Kulikov_Castem_S13a4.txt};
        \addplot file {Images/m15p15_Kulikov_Castem_S13a10.txt};
        \addplot file {Images/m15p15_Kulikov_Castem_S13a100.txt};
        \nextgroupplot[width=7.5cm,height=7.5cm,xlabel={$\overline{\sigma}_{23}(0,b/2)$},
          xmin=-0.5, xmax=0.5, ymin=-0.5, ymax=0.5,
        ]
        \addplot file {Images/m15p15_Kulikov_S23a2.txt};
        \addplot file {Images/m15p15_Kulikov_S23a4.txt};
        \addplot file {Images/m15p15_Kulikov_S23a10.txt};
        \addplot file {Images/m15p15_Kulikov_S23a100.txt};
        \addplot file {Images/m15p15_Kulikov_Castem_S23a2.txt};
        \addplot file {Images/m15p15_Kulikov_Castem_S23a4.txt};
        \addplot file {Images/m15p15_Kulikov_Castem_S23a10.txt};
        \addplot file {Images/m15p15_Kulikov_Castem_S23a100.txt};
        \nextgroupplot[width=7.5cm,height=7.5cm,xlabel={$\overline{\sigma}_{23}(a/2,0)$},
          xmin=-1.2, xmax=0,ymin=-0.5,ymax=0.5,
        ]
        \addplot file {Images/m15p15_Kulikov_S23s2.txt};
        \addplot file {Images/m15p15_Kulikov_S23s4.txt};
        \addplot file {Images/m15p15_Kulikov_S23s10.txt};
        \addplot file {Images/m15p15_Kulikov_S23s100.txt};
        \addplot file {Images/m15p15_Kulikov_Castem_S23s2.txt};
        \addplot file {Images/m15p15_Kulikov_Castem_S23s4.txt};
        \addplot file {Images/m15p15_Kulikov_Castem_S23s10.txt};
        \addplot file {Images/m15p15_Kulikov_Castem_S23s100.txt};
      \end{groupplot}%
      \node at ($(group c1r2.south)!0.5!(group c2r2.south)+(0,0cm)$) [inner sep=0pt,anchor=north, yshift=-8ex] {\pgfplotslegendfromname{legend1_m15p15}};
    \end{tikzpicture}
  }
  \caption{Distribution of the dimensionless transverse shear through the thickness of the [-15/15] composite plate for various length-to-thickness ratios. For clarity, only one FEM curve per plot is presented, for a different length-to-thickness ratio in each plot.}
  \label{fig:m15p15various}
\end{figure}
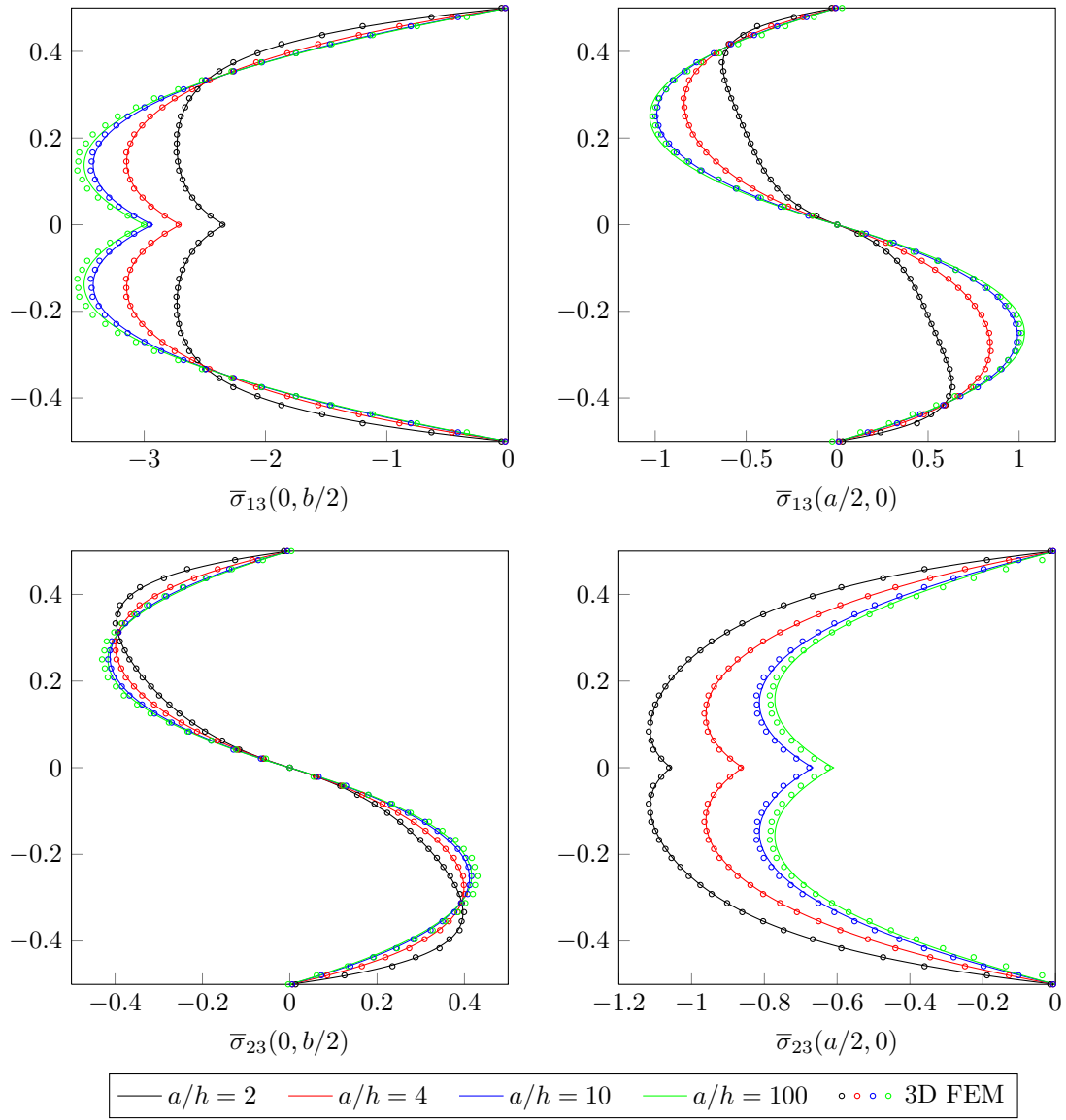

We reproduce here a study which has been done in reference~\cite{Kulikov2012}. A square composite plate with stacking sequence [-15/15] and boundary conditions described in section~\ref{sec:antisymcase} is loaded with a bi-sine load of order $(m,n)=(1,1)$. Four length-to-thickness ratios are considered, $a/h=2,4,10,100$. Figure~\ref{fig:m15p15various} shows the dimensionless transverse shear stresses at points $A(a/2,0)$ and $B(0,b/2)$. Two of these four plots have also been given in the reference~\cite{Kulikov2012}. In addition, finite element computations have been performed with a $24\times 24\times(2\times 12)$ $20$-nodes hexaedron mesh. Values of dimensionless deflection, transverse stresses and fundamental frequency are reported in table~\ref{tab:m15p15_1}. Values from reference~\cite{Kulikov2012} are compared with values from this study, including the analytical ones and those obtained by FEM analysis. Results of the present study agree perfectly with those of reference~\cite{Kulikov2012}, except for a sign on some shear stresses. Table~\ref{tab:m15p15_1} also presents a relative error which is the relative difference between analytical and FEM values. This relative error shows that the particular boundary conditions used in the 3D FEM model are suitable to this type of study. Other set of boundary conditions have been tested but this is the only one that gives so good results.
\iftoggle{submission}{}{\tikzsetnextfilename{m15p15}}

\begin{table*}
	\centering
		\begin{tabular}{c r d{2.6} d{2.6} d{2.6} d{2.6} d{2.6} d{2.6}}
      \toprule
      \multicolumn{1}{c}{$a/h$} & 
			\multicolumn{1}{c}{source} & 
			\multicolumn{1}{c}{$\overline{w}(\tfrac{a}{2},\tfrac{a}{2},0)$} & 
      \multicolumn{1}{c}{$\overline{\sigma}_{13}(\tfrac{a}{2},0,z_1)$} &
			\multicolumn{1}{c}{$\overline{\sigma}_{13}(0,\tfrac{a}{2},z_2)$} &
			\multicolumn{1}{c}{$\overline{\sigma}_{23}(0,\tfrac{a}{2},z_2)$} &  
			\multicolumn{1}{c}{$\overline{\sigma}_{23}(\tfrac{a}{2},0,z_3)$} &			
			\multicolumn{1}{c}{$\overline{\omega}_0$} \\
      \midrule
			  2   &       present           &  4.55480   & 2.71950   & -0.569460  & -0.382364 & 1.10746  & 4.46950  \\
            &       3D FEM			      &  4.55474   & 2.72202   & -0.570113  & -0.382882 & 1.11078  & 4.46952  \\ 
      \multicolumn{1}{c}{} & 
			\multicolumn{1}{c}{\em{rel. err.}} & 
			\multicolumn{1}{c}{\em{-13\ppm}} & 
      \multicolumn{1}{c}{\em{0.93\ppt}} &
			\multicolumn{1}{c}{\em{1.2\ppt}} &
			\multicolumn{1}{c}{\em{1.4\ppt}} &  
			\multicolumn{1}{c}{\em{3.0\ppt}} &			
			\multicolumn{1}{c}{\em{4.5\ppm}} \\
      \midrule
        4   & ref.~\cite{Kulikov2012} &  1.7059    & 3.1447    &  0.84091   &  0.39883  & 0.96037  &          \\  
            &       present           &  1.70585   & 3.14472   & -0.840910  & -0.398830 & 0.960372 & 7.45479  \\
            &       3D FEM			      &  1.70584   & 3.14789   & -0.841323  & -0.399478 & 0.964838 & 7.45482  \\ 
      \multicolumn{1}{c}{} & 
			\multicolumn{1}{c}{\em{rel. err.}} & 
			\multicolumn{1}{c}{\em{-5.9\ppm}} & 
      \multicolumn{1}{c}{\em{1.0\ppt}} &
			\multicolumn{1}{c}{\em{0.49\ppt}} &
			\multicolumn{1}{c}{\em{1.6\ppt}} &  
			\multicolumn{1}{c}{\em{4.7\ppt}} &			
			\multicolumn{1}{c}{\em{4.0\ppm}} \\
      \midrule
      &&&
      \multicolumn{1}{c}{$\overline{\sigma}_{13}(\tfrac{a}{2},0,z_4)$} &
			\multicolumn{1}{c}{$\overline{\sigma}_{13}(0,\tfrac{a}{2},z_5)$} &
			\multicolumn{1}{c}{$\overline{\sigma}_{23}(0,\tfrac{a}{2},z_5)$} &  
			\multicolumn{1}{c}{$\overline{\sigma}_{23}(\tfrac{a}{2},0,z_6)$} &\\			
      \midrule
       10   & ref.~\cite{Kulikov2012} &  0.80272   & 3.4209    &  0.99113  &  0.41169  & 0.81378  &          \\  
            &       present           &  0.802721  & 3.42088   & -0.991131 & -0.411688 & 0.813781 & 11.0265  \\
            &       3D FEM			      &  0.802716  & 3.43544   & -0.995506 & -0.415081 & 0.820914 & 11.0265  \\ 
      \multicolumn{1}{c}{} & 
			\multicolumn{1}{c}{\em{rel. err.}} & 
			\multicolumn{1}{c}{\em{-6.2\ppm}} & 
      \multicolumn{1}{c}{\em{4.3\ppt}} &
			\multicolumn{1}{c}{\em{4.4\ppt}} &
			\multicolumn{1}{c}{\em{8.2\ppt}} &  
			\multicolumn{1}{c}{\em{8.8\ppt}} &			
			\multicolumn{1}{c}{\em{0.0\ppm}} \\
      \midrule
      100   &       present           &  0.622318  &  3.49589  & -1.02916  & -0.416318 & 0.770810 & 12.6740  \\
            &       3D FEM			      &  0.622236  &  3.54552  & -1.01019  & -0.428656 & 0.783020 & 12.6748  \\
      \multicolumn{1}{c}{} & 
			\multicolumn{1}{c}{\em{rel. err.}} & 
			\multicolumn{1}{c}{\em{-130\ppm}} & 
      \multicolumn{1}{c}{\em{14\ppt}} &
			\multicolumn{1}{c}{\em{18\ppt}} &
			\multicolumn{1}{c}{\em{30\ppt}} &  
			\multicolumn{1}{c}{\em{16\ppt}} &			
			\multicolumn{1}{c}{\em{63\ppm}} \\
		\end{tabular}
	\caption{Values of dimensionless deflection and shear stresses for the [-15/15] laminate, with $z_1=0.145$, $z_2=-0.280$, $z_3=0.125$, $z_4=0.140$, $z_5=-0.255$, $z_6=0.150$ for the static load of order $(m,n)=(1,1)$, and values of the fundamental frequency.}
	\label{tab:m15p15_1}
\end{table*}
\subsection{General laminates}
The same study than the one of the [-15/15] laminate is done for two laminates that are neither of cross-ply type nor of antisymmetric angle-ply type. The first of them has a [0/30/0] stacking sequence. It is a symmetric laminate, hence it is not properly speaking a general laminate, but it necessitates the same process than the second laminate which stacking sequence is [-10/0/40]. As described in section~\ref{sec:genecase}, the SSS condition is prescribed for a single order loading ($(m,n)=(1,1)$ in our case) with the help of a supplementary $\cos(\xi x)\cos(\eta y)$ term added to the loading.
\par
\begin{figure}
	\centering
  \iftoggle{submission}{
    \includegraphics{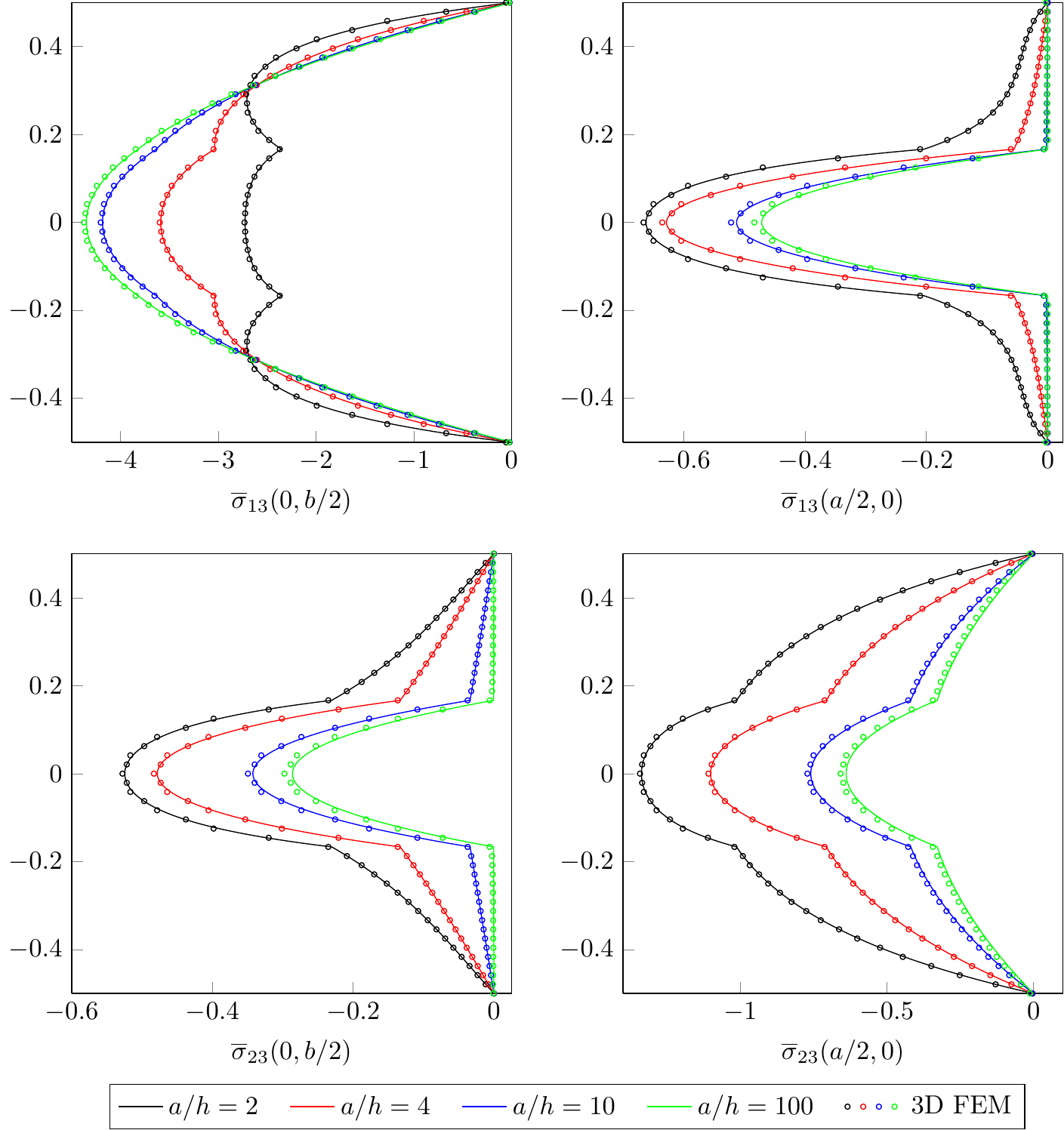}
  }{
    \begin{tikzpicture}
      \begin{groupplot}
        [
            group style=
                {
                group size =2 by 2,
                xlabels at=edge bottom,
                ylabels at=edge left,
                horizontal sep = 1.5 cm,
                vertical sep = 1.5 cm
                },
            width=7.5cm,height=7.5cm,
            tickpos=left,        
            cycle list={{solid, black},{solid, red},{solid, blue},{solid, green},{black, only marks, mark=o},
                        {red, only marks, mark=o},{blue, only marks, mark=o},{green, only marks, mark=o}},
            legend columns=-1,
            transpose legend,
            legend cell align = left,
        ]
        \nextgroupplot[width=7.5cm,height=7.5cm,xlabel={$\overline{\sigma}_{13}(0,b/2)$},
          xmin=-4.5, xmax=0,ymin=-0.5,ymax=0.5,
          legend to name = legend1_p0p30p0
        ]
        \addplot file {Images/p0p30p0_S13s2.txt};
        \addplot file {Images/p0p30p0_S13s4.txt};
        \addplot file {Images/p0p30p0_S13s10.txt};
        \addplot file {Images/p0p30p0_S13s100.txt};
        \addplot file {Images/p0p30p0_Castem_S13s2.txt};
        \addplot file {Images/p0p30p0_Castem_S13s4.txt};
        \addplot file {Images/p0p30p0_Castem_S13s10.txt};
        \addplot file {Images/p0p30p0_Castem_S13s100.txt};
        \legend{$a/h=2\quad$, $a/h=4\quad$, $a/h=10\quad$, $a/h=100\quad$, $$, $$, $$, ~3D FEM}
        \nextgroupplot[width=7.5cm,height=7.5cm,xlabel={$\overline{\sigma}_{13}(a/2,0)$},
          xmin=-0.7, xmax=0.025, ymin=-0.5, ymax=0.5,
        ]
        \addplot file {Images/p0p30p0_S13a2.txt};
        \addplot file {Images/p0p30p0_S13a4.txt};
        \addplot file {Images/p0p30p0_S13a10.txt};
        \addplot file {Images/p0p30p0_S13a100.txt};
        \addplot file {Images/p0p30p0_Castem_S13a2.txt};
        \addplot file {Images/p0p30p0_Castem_S13a4.txt};
        \addplot file {Images/p0p30p0_Castem_S13a10.txt};
        \addplot file {Images/p0p30p0_Castem_S13a100.txt};
        \nextgroupplot[width=7.5cm,height=7.5cm,xlabel={$\overline{\sigma}_{23}(0,b/2)$},
          xmin=-0.6, xmax=0.025, ymin=-0.5, ymax=0.5,
        ]
        \addplot file {Images/p0p30p0_S23a2.txt};
        \addplot file {Images/p0p30p0_S23a4.txt};
        \addplot file {Images/p0p30p0_S23a10.txt};
        \addplot file {Images/p0p30p0_S23a100.txt};
        \addplot file {Images/p0p30p0_Castem_S23a2.txt};
        \addplot file {Images/p0p30p0_Castem_S23a4.txt};
        \addplot file {Images/p0p30p0_Castem_S23a10.txt};
        \addplot file {Images/p0p30p0_Castem_S23a100.txt};
        \nextgroupplot[width=7.5cm,height=7.5cm,xlabel={$\overline{\sigma}_{23}(a/2,0)$},
          xmin=-1.4, xmax=0.1,ymin=-0.5,ymax=0.5,
        ]
        \addplot file {Images/p0p30p0_S23s2.txt};
        \addplot file {Images/p0p30p0_S23s4.txt};
        \addplot file {Images/p0p30p0_S23s10.txt};
        \addplot file {Images/p0p30p0_S23s100.txt};
        \addplot file {Images/p0p30p0_Castem_S23s2.txt};
        \addplot file {Images/p0p30p0_Castem_S23s4.txt};
        \addplot file {Images/p0p30p0_Castem_S23s10.txt};
        \addplot file {Images/p0p30p0_Castem_S23s100.txt};
      \end{groupplot}%
      \node at ($(group c1r2.south)!0.5!(group c2r2.south)+(0,0cm)$) [inner sep=0pt,anchor=north, yshift=-8ex] {\pgfplotslegendfromname{legend1_p0p30p0}};
    \end{tikzpicture}
  }
  \caption{Distribution of the dimensionless transverse shear through the thickness of the [0/30/0] composite plate for various length-to-thickness ratios. For clarity, not all of the FEM curves are plotted.}
  \label{fig:p0p30p0various}
\end{figure}
\begin{figure}
	\centering
  \iftoggle{submission}{
    \includegraphics{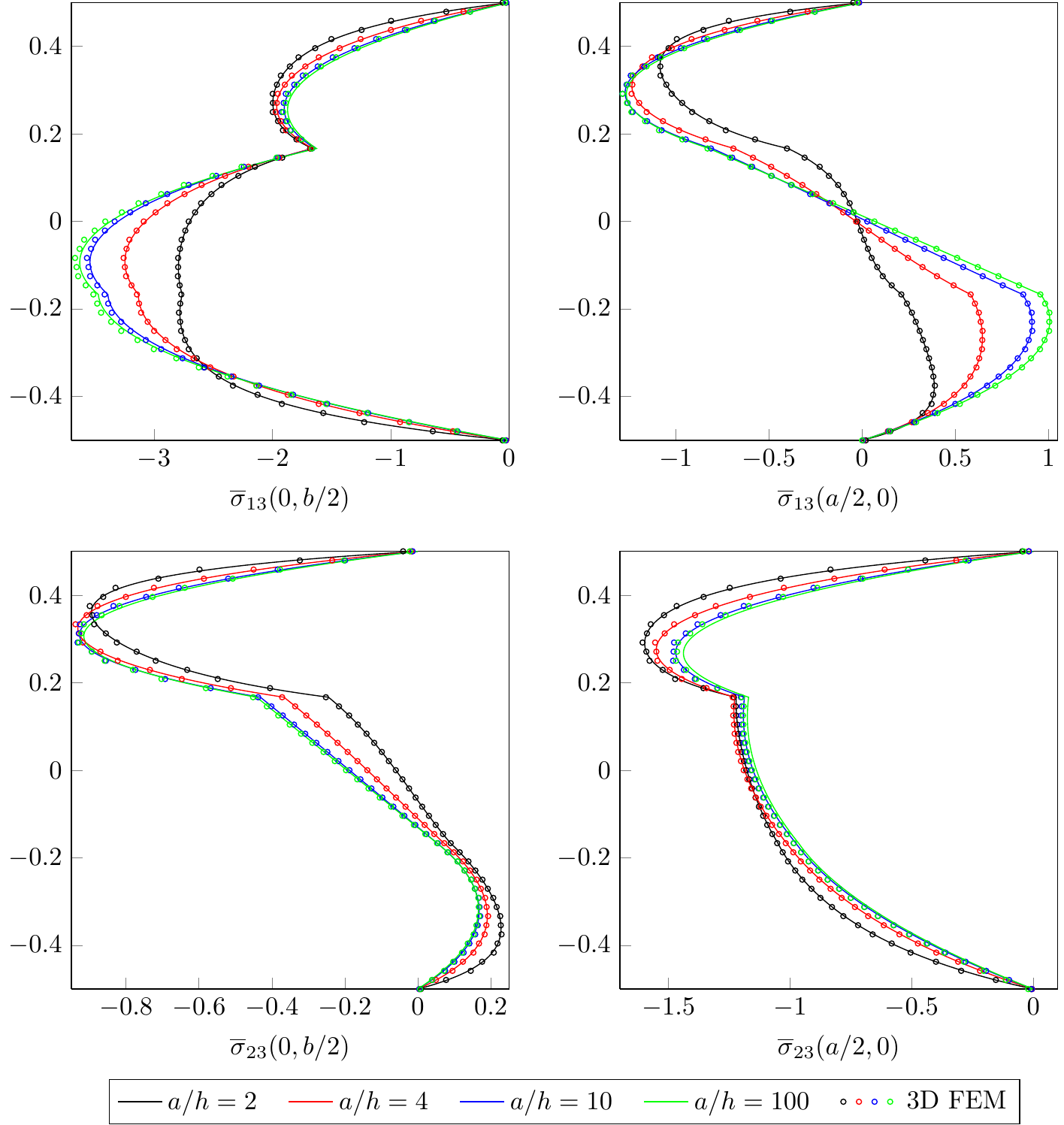}
  }{
    \begin{tikzpicture}
      \begin{groupplot}
        [
            group style=
                {
                group size =2 by 2,
                xlabels at=edge bottom,
                ylabels at=edge left,
                horizontal sep = 1.5 cm,
                vertical sep = 1.5 cm
                },
            width=7.5cm,height=7.5cm,
            tickpos=left,        
            cycle list={{solid, black},{solid, red},{solid, blue},{solid, green},{black, only marks, mark=o},
                        {red, only marks, mark=o},{blue, only marks, mark=o},{green, only marks, mark=o}},
            legend columns=-1,
            transpose legend,
            legend cell align = left,
        ]
        \nextgroupplot[width=7.5cm,height=7.5cm,xlabel={$\overline{\sigma}_{13}(0,b/2)$},
          xmin=-3.7, xmax=0,ymin=-0.5,ymax=0.5,
          legend to name = legend1_m10p0p40
        ]
        \addplot file {Images/m10p0p40_S13s2.txt};
        \addplot file {Images/m10p0p40_S13s4.txt};
        \addplot file {Images/m10p0p40_S13s10.txt};
        \addplot file {Images/m10p0p40_S13s100.txt};
        \addplot file {Images/m10p0p40_Castem_S13s2.txt};
        \addplot file {Images/m10p0p40_Castem_S13s4.txt};
        \addplot file {Images/m10p0p40_Castem_S13s10.txt};
        \addplot file {Images/m10p0p40_Castem_S13s100.txt};
        \legend{$a/h=2\quad$, $a/h=4\quad$, $a/h=10\quad$, $a/h=100\quad$, $$, $$, $$, ~3D FEM}
        \nextgroupplot[width=7.5cm,height=7.5cm,xlabel={$\overline{\sigma}_{13}(a/2,0)$},
          xmin=-1.3, xmax=1.05, ymin=-0.5, ymax=0.5,
        ]
        \addplot file {Images/m10p0p40_S13a2.txt};
        \addplot file {Images/m10p0p40_S13a4.txt};
        \addplot file {Images/m10p0p40_S13a10.txt};
        \addplot file {Images/m10p0p40_S13a100.txt};
        \addplot file {Images/m10p0p40_Castem_S13a2.txt};
        \addplot file {Images/m10p0p40_Castem_S13a4.txt};
        \addplot file {Images/m10p0p40_Castem_S13a10.txt};
        \addplot file {Images/m10p0p40_Castem_S13a100.txt};
        \nextgroupplot[width=7.5cm,height=7.5cm,xlabel={$\overline{\sigma}_{23}(0,b/2)$},
          xmin=-0.95, xmax=0.25, ymin=-0.5, ymax=0.5,
        ]
        \addplot file {Images/m10p0p40_S23a2.txt};
        \addplot file {Images/m10p0p40_S23a4.txt};
        \addplot file {Images/m10p0p40_S23a10.txt};
        \addplot file {Images/m10p0p40_S23a100.txt};
        \addplot file {Images/m10p0p40_Castem_S23a2.txt};
        \addplot file {Images/m10p0p40_Castem_S23a4.txt};
        \addplot file {Images/m10p0p40_Castem_S23a10.txt};
        \addplot file {Images/m10p0p40_Castem_S23a100.txt};
        \nextgroupplot[width=7.5cm,height=7.5cm,xlabel={$\overline{\sigma}_{23}(a/2,0)$},
          xmin=-1.7, xmax=0.1,ymin=-0.5,ymax=0.5,
        ]
        \addplot file {Images/m10p0p40_S23s2.txt};
        \addplot file {Images/m10p0p40_S23s4.txt};
        \addplot file {Images/m10p0p40_S23s10.txt};
        \addplot file {Images/m10p0p40_S23s100.txt};
        \addplot file {Images/m10p0p40_Castem_S23s2.txt};
        \addplot file {Images/m10p0p40_Castem_S23s4.txt};
        \addplot file {Images/m10p0p40_Castem_S23s10.txt};
        \addplot file {Images/m10p0p40_Castem_S23s100.txt};
      \end{groupplot}%
      \node at ($(group c1r2.south)!0.5!(group c2r2.south)+(0,0cm)$) [inner sep=0pt,anchor=north, yshift=-8ex] {\pgfplotslegendfromname{legend1_m10p0p40}};
    \end{tikzpicture}
  }
  \caption{Distribution of the dimensionless transverse shear through the thickness of the [-10/0/40] composite plate for various length-to-thickness ratios. For clarity, not all of the FEM curves are plotted.}
  \label{fig:m10p0p40various}
\end{figure}

Deflections and fundamental frequencies for the [0/30/0] laminate with various length-to-thickness ratios can be seen in table~\ref{tab:p0p30p0} and transverse shear stresses are plotted in figure~\ref{fig:p0p30p0various}. For the [-10/0/40] laminate, results can be seen in table~\ref{tab:m10p0p40} and figure~\ref{fig:m10p0p40various}. The comparison with finite element computations shows that the boundary conditions are adequate. Hence it is possible to have an exact solution for static and dynamic problems even for the most general stacking sequences, but with special boundary conditions.
\iftoggle{submission}{}{\tikzsetnextfilename{p0p30p0}}
\begin{table}
\centering
  \begin{tabular}{c r d{2.6} d{2.6} d{2.6} d{2.6} d{2.6}}
    \toprule
    \multicolumn{1}{c}{$a/h$} &
    \multicolumn{1}{c}{source} &
    \multicolumn{1}{c}{$\overline{w}_{GSS}(\tfrac{a}{2},\tfrac{a}{2},0)$} &
    \multicolumn{1}{c}{$\overline{w}_{SSS}(\tfrac{a}{2},\tfrac{a}{2},0)$} &
    \multicolumn{1}{c}{$p$} &
    \multicolumn{1}{c}{$\overline{\omega}^1_0$} &
    \multicolumn{1}{c}{$\overline{\omega}^2_0$} \\
    \midrule
        2   &       present          &  4.76247  & 4.67167 & -0.138083  &  4.13153 &  4.65351 \\
            &       3D FEM           &  4.76241  & 4.67160 & -0.138084  &  4.13154 &  4.65354 \\
      \multicolumn{1}{c}{} & 
      \multicolumn{1}{c}{\em{rel. err.}} &
      \multicolumn{1}{c}{\em{-13\ppm}} &
      \multicolumn{1}{c}{\em{-15\ppm}} &
      \multicolumn{1}{c}{} &
      \multicolumn{1}{c}{\em{2.4\ppm}} &
      \multicolumn{1}{c}{\em{6.4\ppm}} \\
    \midrule
        4   &       present          &  1.72351  & 1.70511  & -0.103337  &  7.13721 &  7.83090 \\
            &       3D FEM           &  1.72350  & 1.70510  & -0.103337  &  7.13722 &  7.83091 \\
      \multicolumn{1}{c}{} & 
      \multicolumn{1}{c}{\em{rel. err.}} &
      \multicolumn{1}{c}{\em{-5.8\ppm}} &
      \multicolumn{1}{c}{\em{-5.9\ppm}} &
      \multicolumn{1}{c}{} &
      \multicolumn{1}{c}{\em{1.4\ppm}} &
      \multicolumn{1}{c}{\em{1.3\ppm}} \\
    \midrule
       10   &       present          &  0.656135 & 0.653369 & -0.0649254 &  11.8979 &  12.6786 \\
            &       3D FEM           &  0.656132 & 0.653366 & -0.0649256 &  11.8979 &  12.6786 \\
      \multicolumn{1}{c}{} & 
      \multicolumn{1}{c}{\em{rel. err.}} &
      \multicolumn{1}{c}{\em{-4.6\ppm}} &
      \multicolumn{1}{c}{\em{-4.6\ppm}} &
      \multicolumn{1}{c}{} &
      \multicolumn{1}{c}{\em{0.0\ppm}} &
      \multicolumn{1}{c}{\em{0.0\ppm}} \\
    \midrule
      100   &       present          &  0.425176 & 0.423984 & -0.0529466 &  14.9442 &  15.7575 \\
            &       3D FEM           &  0.425135 & 0.423944 & -0.0529420 &  14.9448 &  15.7582 \\
      \multicolumn{1}{c}{} & 
      \multicolumn{1}{c}{\em{rel. err.}} &
      \multicolumn{1}{c}{\em{-96\ppm}} &
      \multicolumn{1}{c}{\em{-94\ppm}} &
      \multicolumn{1}{c}{} &
      \multicolumn{1}{c}{\em{40\ppm}} &
      \multicolumn{1}{c}{\em{44\ppm}} 
  \end{tabular}
  \caption{Values for the [0/30/0] laminate.}
  \label{tab:p0p30p0}
\end{table}
\iftoggle{submission}{}{\tikzsetnextfilename{m10p0p40}}
\begin{table}
\centering
  \begin{tabular}{c r d{2.6} d{2.6} d{2.6} d{2.6} d{2.6}}
    \toprule
    \multicolumn{1}{c}{$a/h$} &
    \multicolumn{1}{c}{source} &
    \multicolumn{1}{c}{$\overline{w}_{GSS}(\tfrac{a}{2},\tfrac{a}{2},0)$} &
    \multicolumn{1}{c}{$\overline{w}_{SSS}(\tfrac{a}{2},\tfrac{a}{2},0)$} &
    \multicolumn{1}{c}{$p$} &
    \multicolumn{1}{c}{$\overline{\omega}^1_0$} &
    \multicolumn{1}{c}{$\overline{\omega}^2_0$} \\
    \midrule
        2   &       present          &  4.64839  & 4.57996  & -0.121332 &  4.18178 &  4.70179 \\
            &       3D FEM           &  4.64831  & 4.57987  & -0.121337 &  4.18180 &  4.70183 \\
      \multicolumn{1}{c}{} & 
      \multicolumn{1}{c}{\em{rel. err.}} &
      \multicolumn{1}{c}{\em{-17\ppm}} &
      \multicolumn{1}{c}{\em{-20\ppm}} &
      \multicolumn{1}{c}{} &
      \multicolumn{1}{c}{\em{4.8\ppm}} &
      \multicolumn{1}{c}{\em{8.5\ppm}} \\
    \midrule
        4   &       present          &  1.79519  & 1.76831  & -0.122380 &  6.87796 &  7.66547 \\
            &       3D FEM           &  1.79518  & 1.76829  & -0.122381 &  6.87797 &  7.66549 \\
      \multicolumn{1}{c}{} & 
      \multicolumn{1}{c}{\em{rel. err.}} &
      \multicolumn{1}{c}{\em{-5.6\ppm}} &
      \multicolumn{1}{c}{\em{-11\ppm}} &
      \multicolumn{1}{c}{} &
      \multicolumn{1}{c}{\em{1.5\ppm}} &
      \multicolumn{1}{c}{\em{2.6\ppm}} \\
    \midrule
       10   &       present          &  0.907571 & 0.897511 & -0.105286 &  9.89646 &  10.8833 \\
            &       3D FEM           &  0.907566 & 0.897505 & -0.105286 &  9.89648 &  10.8833 \\
      \multicolumn{1}{c}{} & 
      \multicolumn{1}{c}{\em{rel. err.}} &
      \multicolumn{1}{c}{\em{-4.4\ppm}} &
      \multicolumn{1}{c}{\em{-5.6\ppm}} &
      \multicolumn{1}{c}{} &
      \multicolumn{1}{c}{\em{2.0\ppm}} &
      \multicolumn{1}{c}{\em{0.0\ppm}} \\
    \midrule
      100   &       present          &  0.732640 & 0.726210 & -0.0936478 &  11.1703 &  12.2682 \\
            &       3D FEM           &  0.732526 & 0.726103 & -0.0936345 &  11.1711 &  12.2690 \\
      \multicolumn{1}{c}{} & 
      \multicolumn{1}{c}{\em{rel. err.}} &
      \multicolumn{1}{c}{\em{-156\ppm}} &
      \multicolumn{1}{c}{\em{-147\ppm}} &
      \multicolumn{1}{c}{} &
      \multicolumn{1}{c}{\em{72\ppm}} &
      \multicolumn{1}{c}{\em{65\ppm}} \\
  \end{tabular}
  \caption{Values for the [-10/0/40] laminate.}
  \label{tab:m10p0p40}
\end{table}

%% file: 05_Conclusion.tex
\section{Conclusion}
The state-space method has been written in a general form that permits to obtain 3D exact solutions for static deflection, and for undamped and damped harmonic responses of laminates. After the choice of trigonometric basis functions for the in-plane variations of displacements $u$, $v$, $w$ and stress components $\sigma_{13}$, $\sigma_{23}$, $\sigma_{33}$, the 3D equilibrium equations are reformulated as a $12\times 12$ first order differential system of equations with respect of the $z$ coordinate. Using the exponentiation of the corresponding $12\times 12$ matrix permits to obtain the $12$ functions that describe the variation through the thickness of these quantities, leading to the complete knowledge of all the fields.
\par
This process permits to treat both class of laminates which are generally considered separately in such studies: cross-ply and anti-symmetrical angle-ply laminates. Further, the same process has been applied to general lamination sequences, and it has been shown that 3D solutions exist for the most general cases, but for special boundary conditions or for special loadings. It is possible to formulate simply supported boundary condition that works for the cross-ply, the anti symmetrical angle-ply and the general classes of laminates. However, for the last case, a deflection of the corners of the plate occurs. It is possible to avoid this deflection if an additional term is added to the loading. Both approaches have been considered for static studies. For the harmonic response, the additional term has not been considered. It has been shown that, for each order of the loading, it exists two natural frequencies with associated mode shapes.
\par
The use of complex stiffnesses lead, with no supplementary effort, to solutions of the damped harmonic response of laminates. Although no values have been given in this paper, it is easy, from there, to obtain exact values for the dissipated power and the loss factor for a given load and a given frequency.
\par
Most of the results of this study have been compared to finite element simulations. The specific boundary conditions that are applied have been presented and discussed. The presented results, especially the plots describing the variations of quantities through the thickness leave no doubt about the pertinence of applied boundary conditions, and about the exactness of the given solutions. 

%% file: 06_Appendix.tex
\section{\texorpdfstring{Matrix $\mathbf{A}^{\ell}$}{Matrix A}}\label{app:app1}
For a given layer $\ell$, the matrix $\mathbf{A}^{\ell}$ of equation~\eqref{eq:solution} is of the form:
\begin{equation}
  \mathbf{A}^{\ell}
  =
  \begin{bmatrix}
      \mathbf{A_1^{\ell}}  &  \mathbf{A_2^{\ell}}  \\\noalign{\medskip}
      \mathbf{A_3^{\ell}}  &  \mathbf{A_4^{\ell}} 
  \end{bmatrix}\\
\end{equation}
where:
\begin{equation}
  \mathbf{A_1^{\ell}}
  =
  \begin{bmatrix}
    0  &  0  &  -\xi  &  4S_{1313}^{\ell}  &      0    &  0  \\\noalign{\medskip}
    0  &  0  & -\eta  &       0     & 4S_{2323}^{\ell} &  0  \\\noalign{\medskip}
    \frac {C_{1133}^{\ell}}{C_{3333}^{\ell}} \xi & \frac {C_{2233}^{\ell}}{C_{3333}^{\ell}} \eta & 0 & 0 & 0 & \frac{1}{C_{3333}^{\ell}} \\\noalign{\medskip}
    \xi^2 Q_{1111}^{\ell}+\eta^2 Q_{1212}^{\ell}-\rho^{\ell} \omega^2 & \eta \xi \left( Q_{1122}^{\ell}+Q_{1212}^{\ell} \right) & 0 & 0 & 0 & -\frac {C_{1133}^{\ell}}{C_{3333}^{\ell}}  \xi \\\noalign{\medskip}
    \eta \xi \left( Q_{1122}^{\ell}+Q_{1212}^{\ell} \right) & \xi^2 Q_{1212}^{\ell}+\eta^2 Q_{2222}^{\ell}-\rho^{\ell} \omega^2 & 0 & 0 & 0 & -\frac {C_{2233}^{\ell}}{C_{3333}^{\ell}} \eta \\\noalign{\medskip}
    0  &  0  &  -\rho^{\ell} \omega^2 &  \xi  &  \eta  &  0
  \end{bmatrix}\\
\end{equation}
\begin{equation}
  \mathbf{A_2^{\ell}}
  =
  \begin{bmatrix}
    0  &  0  &  0  &     0     & 4S_{1323}^{\ell} &  0  \\\noalign{\medskip}
    0  &  0  &  0  & 4S_{1323}^{\ell} &     0     &  0  \\\noalign{\medskip}
    \frac {C_{3312}^{\ell}}{C_{3333}^{\ell}} \eta & \frac {C_{3312}^{\ell}}{C_{3333}^{\ell}} \xi & 0 & 0 & 0 & 0 \\\noalign{\medskip}
    2 \eta Q_{1112}^{\ell} \xi & \xi^2 Q_{1112}^{\ell}+\eta^2 Q_{2212}^{\ell} & 0 & 0 & 0 & \frac {C_{3312}^{\ell}}{C_{3333}^{\ell}} \eta\\\noalign{\medskip}
    \xi^2 Q_{1112}^{\ell}+\eta^2 Q_{2212}^{\ell} & 2 \eta Q_{2212}^{\ell} \xi & 0 & 0 & 0 & \frac {C_{3312}^{\ell}}{C_{3333}^{\ell}}  \xi\\\noalign{\medskip}
    0  &  0  &  0  &  0  &  0  &  0
  \end{bmatrix}
\end{equation}
\begin{equation}
  \mathbf{A_3^{\ell}}
  =
  \begin{bmatrix}
    0  &  0  &  0  &  0  & 4 S_{1323}^{\ell} &  0  \\\noalign{\medskip}
    0  &  0  &  0  & 4 S_{1323}^{\ell} &  0  &  0  \\\noalign{\medskip}
    -\frac {C_{3312}^{\ell}}{C_{3333}^{\ell}} \eta & -\frac {C_{3312}^{\ell}}{C_{3333}^{\ell}} \xi & 0 & 0 & 0 & 0 \\\noalign{\medskip}
    2 \eta Q_{1112}^{\ell} \xi & \xi^2 Q_{1112}^{\ell}+\eta^2 Q_{2212}^{\ell} & 0 & 0 & 0 & -\frac {C_{3312}^{\ell}}{C_{3333}^{\ell}} \eta \\\noalign{\medskip}
    \xi^2 Q_{1112}^{\ell}+\eta^2 Q_{2212}^{\ell} & 2 \eta Q_{2212}^{\ell} \xi & 0 & 0 & 0 & -\frac {C_{3312}^{\ell}}{C_{3333}^{\ell}}  \xi \\\noalign{\medskip}
    0&0&0&0&0&0
\end{bmatrix}
\end{equation}
\begin{equation}
  \mathbf{A_4^{\ell}}
  =
  \begin{bmatrix}
    0  &  0  &  \xi & 4 S_{1313}^{\ell} &  0  &  0  \\\noalign{\medskip}
    0  &  0  & \eta &  0  & 4 S_{2323}^{\ell} &  0  \\\noalign{\medskip}
    -\frac {C_{1133}^{\ell}}{C_{3333}^{\ell}} \xi & -\frac {C_{2233}^{\ell}}{C_{3333}^{\ell}} \eta & 0 & 0 & 0 & \frac{1}{C_{3333}^{\ell}} \\\noalign{\medskip}
    \xi^2 Q_{1111}^{\ell}+\eta^2 Q_{1212}^{\ell}-\rho^{\ell} \omega^2 & \eta \xi \left( Q_{1122}^{\ell}+Q_{1212}^{\ell} \right) & 0 & 0 & 0 & \frac {C_{1133}^{\ell}}{C_{3333}^{\ell}}  \xi\\\noalign{\medskip}
    \eta \xi \left( Q_{1122}^{\ell}+Q_{1212}^{\ell} \right) & \xi^2 Q_{1212}^{\ell}+\eta^2 Q_{2222}^{\ell}-\rho^{\ell} \omega^2 & 0 & 0 & 0 & \frac {C_{2233}^{\ell}}{C_{3333}^{\ell}} \eta\\\noalign{\medskip}
    0  &  0  & -\rho^{\ell} \omega^2 & -\xi & -\eta &  0
\end{bmatrix}
\end{equation}
